\PassOptionsToPackage{hyphens}{url}
\documentclass[aps,pra,onecolumn,superscriptaddress]{revtex4-2}

\usepackage{graphicx}
\usepackage{dcolumn}
\usepackage{pbox}
\usepackage{bm}
\usepackage{amssymb}
\usepackage{amsmath}
\usepackage{siunitx}
\usepackage{wrapfig}
\usepackage{xcolor}
\usepackage{orcidlink}

\hypersetup{pdfpagemode=UseNone}

\begin{document}

\title{Fundamental physics with quantum sensors in space}

\author{Jason Arakawa}
\affiliation{Department of Physics and Astronomy, University of Delaware, Newark, Delaware 19716, USA}

\author{Daniel J. Blumenthal}
\affiliation{Department of Electrical and Computer Engineering, University of California, Santa Barbara, California 93106, USA}

\author{Oliver Buchmueller}
\email{oliver.buchmuller@cern.ch}
\affiliation{Physics Department, Blackett Laboratory, Imperial College London, London, UK}
\affiliation{Physics Department, Oxford University, Oxford, UK}

\author{Dmitry Budker\orcidlink{0000-0002-7356-4814}}
\affiliation{Johannes Gutenberg University, Mainz 55128, Germany}
\affiliation{Helmholtz-Institute, GSI Helmholtzzentrum f\"ur Schwerionenforschung, Mainz 55128, Germany}
\affiliation{Department of Physics, University of California, Berkeley, California 94720, USA}

\author{Luigi Cacciapuoti}
\affiliation{European Space Agency, Keplerlaan 1, Noordwijk, 2201 AZ, NL}

\author{Emily D. Caldwell}
\email{emily.caldwell@nist.gov}
\affiliation{Spectrum Technology and Research Division, National Institute of Standards and Technology, Boulder, Colorado, 80305, USA}

\author{Andrei Derevianko}
\affiliation{Department of Physics, University of Nevada, Reno, Nevada 89557, USA}

\author{Benjamin P. Dix-Matthews}
\affiliation{International Centre for Radio Astronomy Research, The University of Western Australia, Crawley, WA 6009, Australia}

\author{Michael A. Fedderke}
\affiliation{Perimeter Institute for Theoretical Physics, 31 Caroline St N, Waterloo, ON, N2L 2Y5, Canada}

\author{Naceur Gaaloul}
\affiliation{Institut für Quantenoptik, Leibniz Universität Hannover, Welfengarten 1, 30167 Hannover, Germany}

\author{Susan Gardner}
\affiliation{Department of Physics and Astronomy, University of Kentucky, 
Lexington, KY 40506-0055, USA} 

\author{Jonah Glick}
\affiliation{Department of Physics and Astronomy and Center for Fundamental Physics, Northwestern University, Evanston, Illinois, USA}
\affiliation{Fermi National Accelerator Laboratory, Batavia, Illinois, USA}

\author{Paul Hamilton}
\email{paul.hamilton@ucla.edu}
\affiliation{Department of Physics and Astronomy, University of California, Los Angeles, California, 90095, USA}

\author{Elena Jordan}
\affiliation{Physikalisch-Technische Bundesanstalt (PTB), Bundesallee 100, 38116 Braunschweig, Germany}

\author{Hidetoshi Katori}
\affiliation{Department of Applied Physics, The University of Tokyo, 7-3-1 Hongo, Bunkyo-ku, Tokyo 113-8656, Japan}
\affiliation{Spacetime Engineering Team, RIKEN Center for Advanced Photonics, 2-1 Hirosawa, Wako, Saitama 351-0198, Japan}

\author{Peter Kurczynski}
\affiliation{NASA Goddard Space Flight Center, Greenbelt, MD 20771, USA}

\author{David R. Leibrandt\orcidlink{0000-0003-1452-4879}}
\email{leibrandt@ucla.edu}
\affiliation{Department of Physics and Astronomy, University of California, Los Angeles, California, 90095, USA}

\author{Anna L. Marchant}
\affiliation{Rutherford Appleton Laboratory, UKRI--STFC, Harwell Campus, Didcot, OX11 0QX, UK}

\author{Indra Periwal}
\affiliation{Department of Physics, Stanford University, Stanford, California, 94305, USA}

\author{Igor Pikovski}
\affiliation{Department of Physics, Stevens Institute of Technology, Hoboken, New Jersey, 07030, USA}
\affiliation{Department of Physics, Stockholm University, SE-106 91 Stockholm, Sweden}

\author{Marianna S. Safronova}
\affiliation{Department of Physics and Astronomy, University of Delaware, Newark, Delaware 19716, USA}

\author{Christian Sanner}
\affiliation{Department of Physics, Colorado State University, Fort Collins, CO 80523, USA}

\author{Sascha W. Schediwy}
\affiliation{International Centre for Radio Astronomy Research, The University of Western Australia, Crawley, WA 6009, Australia}

\author{Volodymyr Takhistov}   
\affiliation{International Center for Quantum-field Measurement Systems for Studies of the Universe and Particles (QUP, WPI), High Energy Accelerator Research Organization (KEK), Oho 1-1, Tsukuba, Ibaraki 305-0801, Japan}
\affiliation{Theory Center, Institute of Particle and Nuclear Studies (IPNS), High Energy Accelerator Research Organization (KEK), Tsukuba 305-0801, Japan}
\affiliation{Graduate University for Advanced Studies (SOKENDAI), 1-1 Oho, Tsukuba, Ibaraki 305-0801, Japan}
\affiliation{Kavli Institute for the Physics and Mathematics of the Universe (WPI), UTIAS, The University of Tokyo, Kashiwa, Chiba 277-8583, Japan}

\author{Yu-Cheng Tien}
\affiliation{Department of Physics, National Taiwan University, Taipei 10617, Taiwan}

\author{Yu-Dai Tsai}
\affiliation{School of Mathematical and Physical Sciences, University of Sheffield, Sheffield S3 7RH, UK}

\author{Jason R. Williams}
\affiliation{Jet Propulsion Laboratory, California Institute of Technology, Pasadena, California, 91109, USA}

\date{November 16, 2026}

\begin{abstract}

Quantum sensors such as atomic clocks, atom interferometers, and atomic
magnetometers can be used as sensitive probes of physics beyond the Standard
Model. Space offers microgravity, long baselines, low environmental noise,
access to varying gravitational potentials, and unique opportunities involving
the Sun, Moon, planets, and asteroids. These proceedings of the Aspen Center
for Physics workshop on
``Fundamental physics in space with quantum sensors,'' held February
22--27, 2026, summarize the scientific case, current capabilities, and space
mission concepts for investigating dark matter, testing general relativity,
and detecting gravitational waves in frequency bands inaccessible from the
ground~\cite{WorkshopWebsite2026}.
\end{abstract}

\maketitle

\section{Introduction}

The Standard Model of particle physics and Einstein's general
relativity (GR) describe nature with remarkable accuracy. The ordinary
matter described by the Standard Model, however, accounts for only about
$5\%$ of the energy content of the universe.
The remaining $95\%$, roughly $27\%$ in the form of cold dark
matter and $68\%$ in the form of dark energy, have been inferred
from a wide range of astronomical observations, but have not been
identified at the microscopic level. Decades of searches for weakly
interacting massive particles (WIMPs), motivated in part by models of
new physics at the electroweak scale, have not established a dark-matter
signal, although a recent LZ analysis reports a single high-energy
nuclear-recoil candidate at the $2.6\sigma$ level~\cite{Akerib2026}.
The field has broadened its attention to dark-matter candidates
spanning an enormous range of masses, from
ultralight bosonic fields lighter than $10^{-22}$~eV up to composite
objects with macroscopic size. Another fundamental unsolved problem is
the tension between GR and quantum mechanics: GR is a classical theory
of dynamical spacetime, while quantum
mechanics is formulated against a fixed background. The two
frameworks together yield a number of well-posed but experimentally unresolved
questions, including the behavior of quantum superpositions in
curved spacetime, the role of proper time for quantum systems, and
the validity of the equivalence principle for delocalized quantum
objects. Gravitational waves offer a complementary route to these
questions. Laboratory tests probe gravity in the weak-field regime;
waves from merging black holes carry the imprint of spacetime at its
most extreme, strongly curved and rapidly changing, and a stochastic
background would carry the imprint of the universe at energies beyond
the reach of any collider. Each new frequency band opens a new
population of sources, and the microhertz and decihertz bands, home to
the most massive black hole binaries in the universe and to possible
early-universe signals, remain unobserved and are inaccessible from
the ground.

Quantum sensors are well matched to these questions because their
sensitivity to small perturbations grows with the time over which a
quantum state can be held coherent, with the size of the experimental
baseline, and with the cleanliness of the environment. All three are
optimized in space.

Earth is a difficult environment for high-precision physics. Seismic
motion driven by ocean waves and tectonic activity, gravity gradients
from atmospheric and hydrological mass redistribution, anthropogenic
vibration, and stray electromagnetic fields all impose noise floors
that are particularly punishing at low frequencies. Even the largest
terrestrial interferometers are limited in size by the curvature of
the planet.

Space provides four advantages that no ground facility can match.
First, free fall lengthens the interrogation time $T$ available to an
atom interferometer from about one second in the tallest terrestrial
fountains to minutes in orbit. Because the phase response of an atom interferometer
scales as
$
\Delta\phi \;\propto\; T^{2},
$
this corresponds to a sensitivity gain of order $10^{4}$. Second,
baselines of millions of kilometers are achievable with formation-flying
spacecraft, opening gravitational-wave frequency bands that are
inaccessible from the ground. Third, suitable space environments and
instrument designs can provide high vacuum, low temperatures, and
isolation from terrestrial seismic noise, supporting long coherence
times and stable operation. Fourth, the solar system itself
becomes a laboratory: the Sun and the giant planets may host
gravitationally bound dark-matter overdensities; eccentric or
solar-approaching orbits sample varying gravitational potentials;
asteroids can serve as extraordinarily quiet test masses; and the
Moon offers surface sites for long-duration experiments, including access
to cryogenic environments in permanently shadowed polar regions.
\section{Workshop summary}

\subsection{Scientific objectives}
\label{sec:objectives}

The workshop focused its scientific discussion on three
questions that quantum sensors in space are positioned to answer: what
lies in the gravitational-wave frequency bands that no existing or
planned detector covers, what dark matter is made of at the
microscopic level, and how gravity acts on quantum systems. The
subsections below take each question in turn, describing the physics
at stake, the measurements that would address it, and the reasons
those measurements require space. Section~\ref{sec:sensors} then
turns to the sensors themselves and translates the three questions
into requirements on clock and interferometer performance. The
corresponding clock-performance requirements and atom-interferometer
sensitivities are summarized in Figs.~\ref{fig:summary-clocks}
and~\ref{fig:summary-AI}, respectively.

\newpage
\subsubsection{What lies in the unexplored gravitational-wave frequency bands?}

Gravitational-wave astronomy currently spans two widely separated
windows. Pulsar timing arrays probe the nanohertz band, where they have
recently found evidence for a stochastic background attributed to the
population of supermassive black hole
binaries~\cite{Agazie2023, Antoniadis2023, Reardon2023, Xu2023}, and
ground-based laser interferometers observe stellar-mass compact binary
coalescences between roughly $10$ and $10^3$\,Hz. The space-based LISA
mission will add a third window from roughly $10^{-4}$ to
$10^{-1}$\,Hz, where massive black hole mergers and galactic binaries
dominate~\cite{LISA:2017pwj}. Two gaps will remain. Between the pulsar
timing and LISA bands lies the microhertz band, roughly $10^{-7}$ to
$10^{-4}$\,Hz, for which no dedicated detector exists or is
planned~\cite{Sesana2019}. Between LISA and the ground-based detectors
lies the decihertz band, roughly $10^{-2}$ to $1$\,Hz, for which the
leading proposals are space-based atom
interferometers~\cite{Sedda:2020vwo, Badurina:2021rgt}. Each gap hosts
a distinct population of sources, and both lie below the frequencies at
which seismic and Newtonian noise become prohibitive on the ground, so
their full exploration requires space-based detectors. The two bands
are discussed in turn below.

The decihertz band hosts mergers of intermediate-mass black holes with masses of
$10^2$--$10^4\,M_\odot$, compact binaries in their final hours to years
before coalescence, and stochastic backgrounds from early-universe phase
transitions~\cite{Sedda:2020vwo, Badurina:2021rgt}. A particularly
compelling feature of this band is the timescale on which sources evolve
through it. A compact binary
approaching merger sweeps to higher frequencies, spending days to months
in this range before entering the LIGO band. This extended observation
window has a direct consequence for sky localization: as the Earth moves
along its orbit, a space-based detector traces a baseline of order one
astronomical unit over the course of a signal, allowing the source
direction to be reconstructed with much higher angular precision than a
snapshot observation permits. For a nearby binary neutron star system,
this translates to a sky localization accurate to arcminutes or better,
hours to days before the merger itself is detectable by ground-based
instruments~\cite{Sedda:2020vwo, Badurina:2021rgt}. The practical
consequence is a qualitative change in how multi-messenger follow-up can
be organized. Electromagnetic and neutrino observatories can be pointed
at a well-defined patch of sky in advance of the merger, rather than
responding to a post-merger alert with a large localization uncertainty.
Events like GW170817, where the electromagnetic counterpart was
identified only after a search across a large sky-localization
region~\cite{Monitor:2017mdv}, illustrate what is currently possible.
The decihertz band would make such detections systematic rather than
fortuitous.

Space-based atom interferometers, operating on satellite baselines of
order $10^4$\,km, are natural instruments for this band. As illustrated
in Fig.~\ref{fig:summary-AI}, concepts along the lines of AEDGE would
achieve competitive strain sensitivities in the decihertz range while
extending sensitivity to frequencies two to three orders of magnitude
below the optimal frequency range of ground-based
detectors~\cite{El-Neaj2020, Badurina:2021rgt}. The terrestrial program
now under construction, including AION and MAGIS-100, together with the
100-m-scale AICE facility proposed at CERN~\cite{AICE:2026}, will
develop and validate the atom optics, noise mitigation, and quantum
enhancement techniques on which the space-based program depends,
while delivering first sensitivity to both gravitational waves and
ultralight dark matter in the decihertz
band~\cite{Badurina2020, Abe2021, Badurina:2021rgt}. Optical clocks
offer a second route into this band. Two drag-free spacecraft, each
carrying an optical clock and exchanging laser light over a baseline of
$10^{4}$--$10^{6}$\,km, register a passing wave as a shift in the
frequency of the received light relative to the local
clock~\cite{Kolkowitz2016, Ganapathy2026}; the two approaches are
closely related, since in a gradiometer the atoms' internal clock
transition likewise serves as the phase reference for a single laser
baseline~\cite{Graham2013}. As indicated in
Fig.~\ref{fig:summary-clocks}, this is the one clock application that
requires instability below $10^{-18}/\sqrt{\tau}$. The CHRONOS clock
concept is described in the contribution by Kurczynski, and the
atom-interferometer program, from terrestrial prototypes to mission
concepts, in the contributions by Buchmueller, Glick, and Periwal in
Sec.~\ref{sec:proceedings}.

The microhertz band hosts the most massive black hole binaries in the
universe. Binaries with total masses of $10^{7}$--$10^{10}\,M_\odot$
evolve through microhertz frequencies: the most massive, above $10^{9}\,M_\odot$, merge within the band, while
lighter binaries spend months to centuries there before entering the
LISA band. The band therefore connects the background seen by pulsar timing arrays to
individually resolvable, evolving binaries that would allow the most
massive black holes in the universe to be studied one at a time. Extreme-mass-ratio inspirals into the most
massive black holes, which are too low in frequency for LISA, also emit
here, as do cosmological sources: first-order phase transitions in the
early universe at temperatures between the QCD and electroweak scales,
cosmic strings, and an inflationary background would all produce
stochastic signals in this band~\cite{Sesana2019}. Even in space, the
free-fall requirements at these frequencies exceed what LISA Pathfinder
demonstrated~\cite{Armano2018}, which motivates the use of
asteroids~\cite{Fedderke2022} or planets~\cite{Bills2022} as naturally
isolated test masses.

Clocks are the natural instrument for this band. A gravitational wave
passing between two free-falling bodies modulates the light-travel time
between them, so a frequency reference that is stable over the wave
period registers the strain as a fractional Doppler shift on a link
between them~\cite{Estabrook1975, Armstrong2006}. Spacecraft
Doppler tracking of the Cassini mission, referenced to a hydrogen maser,
achieved fractional-frequency instability (Allan deviation) of a few
$10^{-15}$ at averaging times around $10^3$\,s~\cite{Asmar2005}
and set upper limits on gravitational waves in the
$10^{-6}$--$10^{-3}$\,Hz band far above the expected
signals~\cite{Armstrong2003}. Optical clocks in
space with $10^{-18}$ fractional-frequency instability at averaging
times of $10^{4}$--$10^{6}$\,s, combined with optical time-frequency
transfer of comparable stability (Sec.~\ref{sec:clocks}), would
improve on the Cassini reference by three orders of magnitude, with a corresponding gain in strain sensitivity at fixed
baseline and frequency. Several concepts build on this premise: clock-referenced
Doppler measurements between spacecraft~\cite{Loeb2015, Vutha2015},
and ranging between asteroids~\cite{Fedderke2022} or planetary
landers~\cite{Bills2022}, whose large masses suppress the
non-gravitational perturbations that limit engineered test masses. In every case the
requirement is on long-term stability rather than absolute accuracy, as
indicated in Fig.~\ref{fig:summary-clocks}: a clock with
$10^{-15}/\sqrt{\tau}$ instability ($\tau$ in seconds) and a $10^{-18}$
floor reaches $10^{-18}$ at $10^{6}$\,s, the period of a microhertz
wave. Because the
strain response of a single baseline falls once the wave period exceeds
the light-travel time across it, baselines of order an astronomical unit
are preferred, and multiple baselines allow clock noise to be separated
from the common signal or removed altogether~\cite{Fedderke2022}. The asteroid concept is developed in
the contribution by Fedderke in Sec.~\ref{sec:proceedings}.

\subsubsection{What is the microscopic nature of dark matter?}

The nature of dark matter remains one of the central open questions in
physics. WIMPs at the electroweak
scale have been a major target for several decades, and
direct-detection, collider, and indirect searches have placed stringent
constraints on benchmark WIMP models~\cite{LZ:2022ufs, XENON:2023syk,
CMS:2024dark}. Ultralight bosons at the opposite end of the mass range
are motivated independently of those results: the QCD axion, which
resolves the strong CP problem; the dilaton and moduli fields generic to
string theory; and the relaxion, which addresses the electroweak
hierarchy problem, all yield dark-matter candidates with masses far
below an electronvolt~\cite{Antypas2022}. For an ultralight boson
constituting all of the dark matter, the existence of small galaxies
implies a characteristic lower mass scale around
$10^{-22}$\,eV~\cite{Hu:2000ke, Hui:2016ltb}. Lighter particles may
remain viable as a subdominant component, with quantitative bounds
depending on abundance, model, and dataset~\cite{Kobayashi2017}.
Throughout this range,
dark matter behaves not as a gas of individual particles but as a
coherently oscillating classical field, with an oscillation frequency
set by the particle mass, $f \sim m_\phi c^2 / 2\pi\hbar$, and a
coherence time of roughly $10^{6}$ oscillation periods set by the
velocity dispersion of the galactic halo~\cite{Graham:2015ifn,
Antypas2022}.

Three classes of ultralight dark-matter candidates are particularly
relevant for quantum sensor searches. Scalar fields, such as dilatons
and moduli, couple to Standard Model fields through terms that shift
fundamental constants, including the fine-structure constant and particle
masses. This produces oscillations in atomic transition frequencies that
atomic clocks are well suited to detect~\cite{Arvanitaki2015, Safronova2018}.
Pseudoscalar fields, including axions and axion-like particles, couple to
fermion spins and photons; the oscillating field can drive spin
precession and produce oscillating electric dipole moments accessible to
atomic magnetometers, nuclear magnetic resonance, and other spin-based
sensors~\cite{Safronova2018, Antypas2022}. Vector fields,
associated with new gauge symmetries such as $B{-}L$, couple to baryon
and lepton number and generate differential accelerations between test
masses of different composition, making atom interferometers in a
gradiometric configuration a natural search tool~\cite{El-Neaj2020,
Badurina:2021rgt}. The two sensor modalities discussed in these proceedings are
complementary across these classes: atomic clocks are particularly
sensitive to scalar couplings that shift transition frequencies, while
atom interferometers probe the differential phase accumulated between
spatially separated atom clouds, providing sensitivity to both scalar
and vector couplings.

Not all dark-matter candidates produce oscillatory signals. Compact
field configurations, including topological defects such as domain
walls and monopoles, Q-balls, boson stars, and dark-matter
clumps~\cite{Derevianko2014, Baum2022}, would pass through a sensor at
the galactic virial velocity of about $300$\,km/s and produce a
transient rather than a persistent signal. A single unexplained
excursion in one instrument is rarely persuasive, but a network of
sensors with known geometry turns such an encounter into a time-ordered
pattern that encodes the size, velocity, direction, and coupling of the
passing object. The Global Positioning System, whose constellation of
microwave clocks forms a $50{,}000$\,km aperture with two decades of
archival data, has already been used in this way: a search of 16 years
of data found no domain-wall events and improved bounds on quadratic
scalar couplings by orders of magnitude~\cite{Roberts2017, Roberts2018}.
The same networks can search for relativistic bursts of ultralight
fields emitted by compact-object mergers, supernovae, or bosenovae,
using gravitational-wave or electromagnetic detections as
triggers~\cite{Dailey2020, Eby2021, Arakawa2026}, extending
multi-messenger astronomy to new fields. Space-based networks of
optical clocks would improve on the GPS example by several orders of
magnitude in clock stability while adding longer baselines and freedom
from screening. Space-based atom-interferometer configurations could
also probe time-dependent gravitational accelerations and gradients
produced by dark-matter structures. These directions are developed in the contributions by
Derevianko, Takhistov, Arakawa, Fedderke, and Gardner in
Sec.~\ref{sec:proceedings}.

The case for space rests on three effects that no terrestrial
experiment can reproduce. First, in some models with quadratic couplings
to matter, the Earth and its atmosphere can suppress the local scalar-field
amplitude. Space-based sensors can alleviate this screening, with the
extent of recovery depending on the model parameters and orbital
altitude~\cite{Stadnik2021, Banerjee2023,
Arakawa2024}. Second, the dark-matter distribution within the solar
system is poorly known, and overdensities bound to the Sun or the giant
planets could exceed the galactic average by orders of
magnitude~\cite{Budker2023}; a spacecraft carrying two clocks of
different species to about $0.1$\,AU would test for a solar halo, probe
the relaxion, and search for variations of fundamental constants
correlated with the solar gravitational potential~\cite{Tsai2023,
Budker2025}. Third, for masses near $10^{-16}$\,eV the de Broglie
wavelength is comparable to the size of the solar system, so a network
of sensors on separate spacecraft sees a single coherent signal, and
baselines far longer than any terrestrial network separate a
dark-matter signal from local perturbations. The sensors themselves
also gain. Extended free fall and the absence of seismic noise extend
the reach of atom interferometers to the lowest masses, and a
space-based instrument along the lines of AEDGE could probe the
coupling of ultralight dark matter to electrons across
$10^{-21}$--$10^{-15}$\,eV at sensitivities well beyond current limits
from torsion balances and terrestrial atomic
clocks~\cite{El-Neaj2020, Badurina:2021rgt}. Clock comparisons between
space and ground cover oscillation frequencies from $10^{-8}$ to
$1$\,Hz, and comparing a space clock against its own reference cavity
extends the reach to $10^{3}$\,Hz (Figs.~\ref{fig:summary-clocks}
and~\ref{fig:summary-AI}). Terrestrial atom interferometers now under
construction will provide the first interferometric reach into this
parameter space and establish the demonstrated sensitivities on which
a space-based program can build~\cite{Badurina2020, Abe2021}.

\subsubsection{How does gravity act on quantum systems?}

General relativity has passed every test to which it has been
put~\cite{will2014confrontation}, yet it remains a classical theory,
despite a century of effort to reconcile it with quantum mechanics. Clocks and atom
interferometers can test three foundations of GR directly: the
gravitational redshift, which asks whether a clock's rate depends on
the gravitational potential exactly as predicted; the universality of
free fall, which asks whether all matter, including delocalized
quantum matter, falls identically; and local Lorentz invariance, which
asks whether the outcome of an experiment depends on its orientation
and velocity. A fourth question, what happens to a quantum
superposition whose branches experience different proper times, has
no classical analog. Space offers each of these tests something the
ground cannot: potential differences of a few parts in $10^{10}$
along an eccentric orbit, minutes of free fall, and an orbital
velocity and rotation that modulate any anisotropy on a timescale
distinct from terrestrial noise.

The gravitational redshift is tested by comparing clocks at different
potentials and expressing any deviation as a parameter $\alpha$ that
vanishes in GR. Gravity Probe A confirmed the redshift at the
$10^{-4}$ level with a hydrogen maser on a suborbital flight in
1976~\cite{Vessot1980}, and two Galileo satellites in unintended
eccentric orbits reached $2.5\times10^{-5}$ with their onboard
microwave clocks~\cite{Delva2018, Herrmann2018}. ACES, now on the ISS,
targets a few parts in $10^{6}$~\cite{Cacciapuoti2024}. Optical clocks
change the picture: on the ground they have resolved the redshift over
height differences from 450\,m~\cite{Takamoto2020GR} down to about a
centimeter~\cite{zheng2023lab} and, within a single atomic sample,
across a millimeter~\cite{bothwell2022resolving}. The FOCOS mission proposal, a strontium lattice
clock on a highly eccentric Earth orbit linked optically to ground
clocks, could reach $10^{-9}$~\cite{FOCOS2022}, and a single-ion clock built
on current technology could reach below $10^{-8}$. The science case and
implementation of clock-based gravitational-redshift tests are
discussed in the contributions by Safronova and Leibrandt.

The universality of free fall is tested by comparing the accelerations
of two bodies of different composition. With classical test masses,
the MICROSCOPE satellite reached $10^{-15}$~\cite{Touboul2022}. Atom
interferometers extend the test to quantum objects: two isotopes or
two species, or a single species prepared in a superposition of two
internal states~\cite{Rosi2017}, are dropped together and their
differential phase is read out with a common laser. The terrestrial
state of the art is $10^{-12}$ with rubidium isotopes in a 10\,m
fountain~\cite{Asenbaum_EP_PhysRevLett.125.191101}. Because the
sensitivity grows as the square of the interrogation time, orbit is
the natural next step, and the STE-QUEST mission proposal calls for a dual-species
interferometer with the long interrogation times available in orbit
to reach $10^{-17}$~\cite{gaaloul2025stequest, struckmann2024platform},
two orders of magnitude beyond MICROSCOPE and with test masses whose
quantum nature is not in doubt. The Cold Atom Lab has already
demonstrated dual-species interferometry in
orbit~\cite{Elliott2023Nature}. STE-QUEST and the Cold Atom Lab are
discussed in the contributions by Gaaloul and Williams.

Local Lorentz invariance underlies both GR and the Standard Model, and
its possible violation is parametrized by the Standard Model
Extension~\cite{Colladay1998, Mattingly2005}. In the electron sector
the most sensitive probes are trapped-ion clock transitions whose
frequency would depend on the orientation of the atomic orbitals
relative to a preferred frame. The Earth's rotation and orbital motion
modulate any such shift, and Michelson--Morley-type experiments with
calcium and ytterbium ions have improved the bounds by several orders
of magnitude over the past decade~\cite{Pruttivarasin2015, Sanner2019,
Dreissen2022}. All of them rely on the Earth's rotation, so a true
signal competes with terrestrial noise at the same sidereal
frequencies. A clock in orbit introduces a new modulation period,
about 90 minutes on the ISS, and a larger boost, and a measurement
synchronized between Earth and orbit isolates a violation through its
unique correlation signature. A ytterbium-ion clock in space operated
in this way could improve the best bounds a hundredfold without an
ultrastable laser, as proposed in the contribution by Sanner.

Perhaps the most fundamental question is how quantum mechanics itself behaves in
curved spacetime. Proper time enters quantum theory through the phase
of a clock's internal state, so a clock placed in a superposition of
two heights accumulates a superposition of proper times; the resulting
loss and revival of interferometric visibility would be a direct
signature of time dilation acting on a quantum degree of
freedom~\cite{zych2011quantum, dimopoulos2008general}, and the same
mechanism decoheres any composite system with internal
energy~\cite{pikovski2015universal}. Optical ion clocks are
approaching the precision needed to reveal these signatures in the
laboratory~\cite{sorci2025quantum}, and networks of entangled clocks
distributed over a range of altitudes would turn them into a probe of
the spacetime metric itself~\cite{komar2014quantum,
borregaard2025testing, covey2025probing}. A related program asks
whether gravity can entangle two massive objects, which would show
that the gravitational field is quantum~\cite{bose2017spin,
marletto2017gravitationally}; it needs massive superpositions held in
free fall for long times, which is what orbit offers. Beyond atoms, a
levitated ferromagnet whose macrospin precesses in the gravitational
field could measure frame dragging and geodetic precession, a
``gravity probe spin'' that gains from the quiet environment of
space~\cite{fadeev2021gravity, kimball2016precessing}. These
directions are developed in the contributions by Pikovski and Budker.

\subsection{Quantum sensors}
\label{sec:sensors}

Quantum sensors based on cold atoms have matured into a new class of
instruments for fundamental physics. Decades of progress in ion
trapping, laser cooling, Bose--Einstein condensation, frequency-comb
spectroscopy, quantum control, and entanglement have been recognized by
Nobel Prizes in 1989, 1997, 2001, 2005, 2012, and 2022, and have
produced atomic samples that are simultaneously cold (down to
picokelvin temperatures), trapped, and precisely controllable~\cite{Safronova2018}.

Optical atomic clocks now achieve fractional-frequency uncertainties
below $10^{-18}$~\cite{Ludlow2015, Aeppli2024, Marshall2025, Lindvall2025,
Zhang2026}, making measurements of time and frequency sensitive to
gravitational redshifts corresponding to centimeter-scale height
differences~\cite{McGrew2018, Takamoto2020GR, bothwell2022resolving,
zheng2023lab}. Atom interferometers exploit the
wave nature of laser-cooled atoms to measure inertial and gravitational
phase shifts with sensitivities that are beginning to constrain
equivalence-principle violations and couplings of ultralight dark matter
to Standard Model fields in their current terrestrial
configurations~\cite{Buchmueller:2023interferometry, Alonso:2022oot}.

Each of the questions in the preceding section places distinct
requirements on the sensors, and Figs.~\ref{fig:summary-clocks}
and~\ref{fig:summary-AI} map the science cases onto required sensor performance.
For clocks, micrometer-level ephemerides require an inaccuracy near
$10^{-17}$; equivalence-principle tests, microhertz gravitational waves,
optical VLBI, and comparisons of ground clocks that are not in common
view require $10^{-18}$ inaccuracy with an instability of order
$10^{-15}/\sqrt{\tau}$; searches for ultralight dark matter are limited
by instability and by the sensitivity of the clock transition to the
coupling rather than by inaccuracy; and only decihertz gravitational-wave
detection demands instability below $10^{-18}/\sqrt{\tau}$. Every
clock-based case at the $10^{-18}$ level also requires free-space
time-frequency transfer of comparable stability between spacecraft and
the ground. For atom interferometers, whose phase sensitivity grows as
the square of the interrogation time, equivalence-principle tests and
the lowest dark-matter masses call for the longest free-fall times,
while decihertz gravitational waves and the dark-matter gradiometer
configuration benefit from satellite baselines of order $10^{4}$\,km.
Beyond the three fundamental physics questions above, geodesy and Earth observation add an
application-driven case in which the absolute, drift-free character of
quantum inertial sensors addresses the low-frequency limitations of the
electrostatic accelerometers flown on current satellite gravity
missions, and in which clock comparisons measure geopotential
differences directly; these applications appear in the contributions by
Williams, Jordan, and Buchmueller.

The two sensor modalities that meet these requirements, optical atomic
clocks (Sec.~\ref{sec:clocks}) and atom interferometers
(Sec.~\ref{sec:AI}), are described below.

\subsubsection{Optical atomic clocks}
\label{sec:clocks}

\begin{figure}[h!]
\begin{center}
\includegraphics[width=1.0\textwidth]{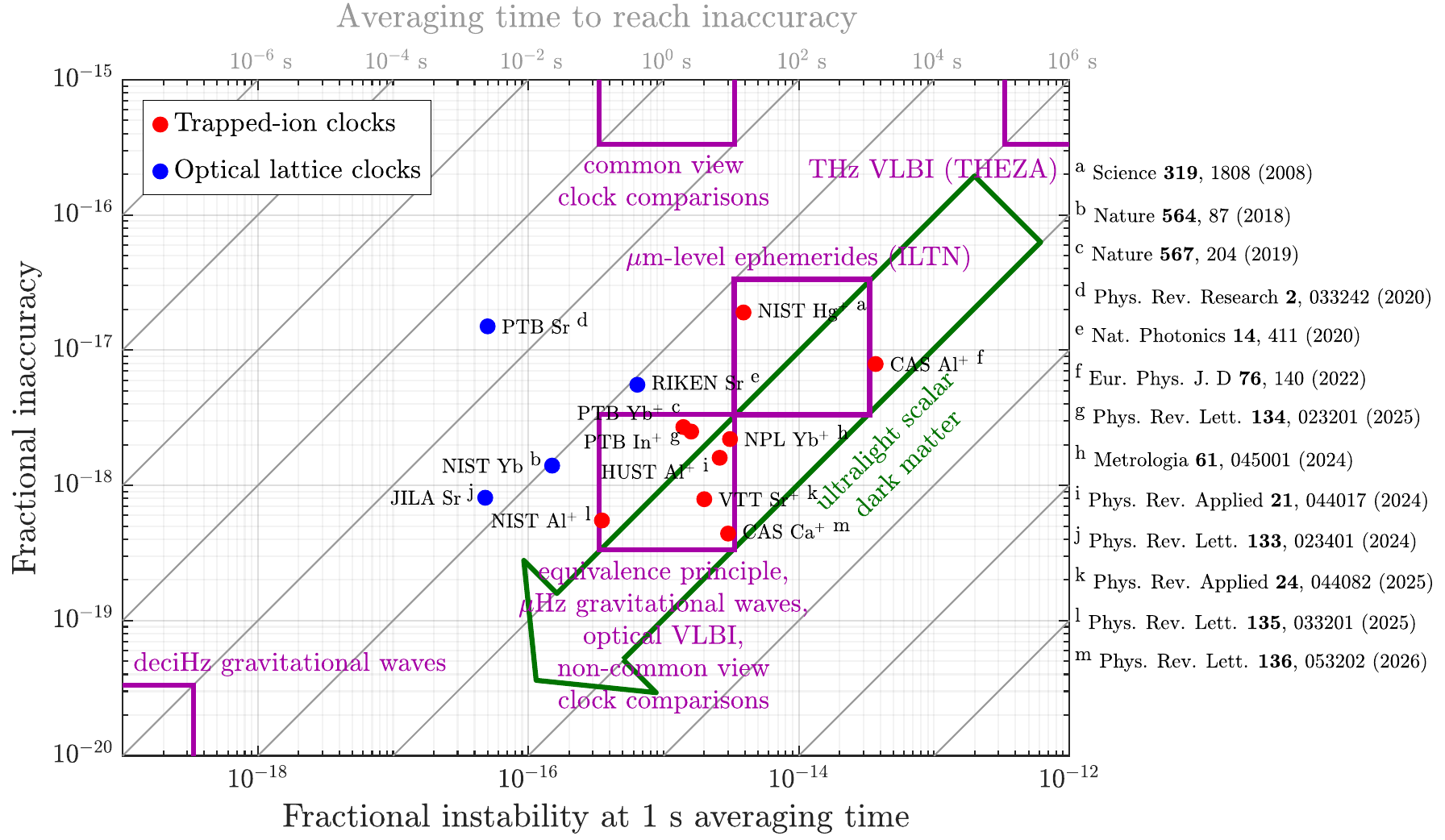}
\caption{\label{fig:summary-clocks}
Performance of laboratory optical clocks compared with the requirements
of the science cases in Sec.~\ref{sec:objectives}. Each point gives the
fractional-frequency instability at 1\,s averaging time and the
fractional inaccuracy of a trapped-ion clock (red) or optical lattice
clock (blue): (a) NIST Hg$^+$~\cite{Rosenband2008}, (b) NIST
Yb~\cite{McGrew2018}, (c) PTB Yb$^+$~\cite{Sanner2019}, (d) PTB
Sr~\cite{Schwarz2020}, (e) RIKEN Sr~\cite{Takamoto2020GR}, (f) CAS
Al$^+$~\cite{Cui2022a}, (g) PTB In$^+$~\cite{Hausser2025}, (h) NPL
Yb$^+$~\cite{Tofful2024}, (i) HUST Al$^+$~\cite{Ma2024}, (j) JILA
Sr~\cite{Aeppli2024}, (k) VTT Sr$^+$~\cite{Lindvall2025}, (l) NIST
Al$^+$~\cite{Marshall2025}, and (m) CAS Ca$^+$~\cite{Zhang2026}.
Diagonal lines show the averaging time at which a clock with white
frequency noise reaches its inaccuracy. Magenta boxes mark the
performance required for common-view clock comparisons, THz
VLBI~\cite{Gurvits2021}, micrometer-level ephemerides~\cite{Bills2022},
tests of the equivalence principle~\cite{FOCOS2022}, $\mu$Hz
gravitational-wave detection~\cite{Fedderke2022}, optical VLBI,
non-common-view comparisons of terrestrial $10^{-18}$ optical clocks, and decihertz gravitational-wave
detection~\cite{Kolkowitz2016, Ganapathy2026}. Green arrows mark
searches for ultralight scalar~\cite{Arvanitaki2015} and
topological~\cite{Derevianko2014} dark matter, whose reach improves
continuously with clock stability and has no fixed inaccuracy
requirement.
}
\end{center}
\end{figure}

Optical atomic clocks are typically based on locking a cavity-stabilized laser to a narrow
electronic transition at hundreds of THz and counting its oscillations
with a frequency comb~\cite{Ludlow2015}. Laser cooling and confinement
in the Lamb--Dicke regime eliminate the first-order Doppler shift and
permit interrogation times of seconds. Optical lattice clocks
interrogate thousands of neutral atoms in a magic-wavelength optical
lattice and achieve the lowest instability, because quantum projection
noise averages down with atom number~\cite{Aeppli2024}; trapped-ion
clocks interrogate one or a few ions in a radio-frequency Paul trap and
achieve the lowest inaccuracy, because the ions are well
isolated from perturbations~\cite{Marshall2025, Lindvall2025,
Zhang2026}. Performance is characterized by inaccuracy, the residual
uncertainty in systematic shifts at long averaging times, and
instability, the statistical uncertainty as a function of averaging
time. Figure~\ref{fig:summary-clocks} places the leading laboratory
clocks on these two axes alongside the requirements of the science
cases.

For space missions, clock performance must be weighed against size, weight, and power
(SWaP) and technology readiness level (TRL). Only warm vapor-cell
optical clocks are near flight readiness: an iodine clock reaches
$5\times10^{-15}$ at one day in 29\,L and 80\,W, has operated at
sea~\cite{Roslund2024}, and is being ruggedized for
space~\cite{VectorAtomic2025, VectorAtomic2025a}, and a two-photon
rubidium clock is in environmental testing for a 2027 launch, with its
frequency comb operating in orbit since March 2026~\cite{QuantX2026}.
Laser-cooled optical clocks remain laboratory or rack-scale commercial
instruments at TRL 2 to 3~\cite{TOPTICA2025, Shimadzu2025}, although
their subsystems have flight heritage: a cavity-stabilized laser has
operated on GRACE Follow-On since 2018~\cite{Abich2019}, and frequency
combs have flown on sounding rockets~\cite{Lezius2016} and in
orbit~\cite{QuantX2026} and are slated for the
ISS~\cite{Kuschewski2024}. Proposed space optical clocks range from a single-ion
clock at $10^{-15}/\sqrt{\tau}$ instability and $10^{-18}$ inaccuracy
in 20\,L and 100\,W to an optical lattice clock at
$10^{-16}/\sqrt{\tau}$ and $10^{-18}$ in about 300\,L and
1\,kW~\cite{FOCOS2022}. The trade space among them is discussed in the
contribution by Leibrandt, and compact clock architectures and photonic
integration in the contributions by Katori and Blumenthal in
Sec.~\ref{sec:proceedings}.

Future $10^{-18}$-level optical clocks in space will require free-space time-frequency transfer techniques that operate with sub-$10^{-18}$ instability and inaccuracy. Currently, two techniques of free-space optical time-frequency transfer have demonstrated instabilities at this level across terrestrial links: two-way optical time transfer with frequency combs~\cite{Caldwell2023,shen2022}, and active phase-stabilization of the link with continuous-wave (CW) lasers~\cite{Gozzard2022, Maron2024}. \citet{Bodine2020} showed that the comb-based technique could compare optical clock offsets through frequency ratio measurements at levels comparable to optical fiber.  

The two primary technical challenges facing the use of these techniques in ground-to-space clock comparisons are the expected high loss and potentially large Doppler shifts. The expected losses for ground-to-space links depend on orbit choice and aperture sizes, and range from 35\,dB to $>$90\,dB for low LEO and GEO, respectively~\cite{Caldwell2023, shen2021}. Comb-based time transfer has been shown to operate through 90\,dB of loss~\cite{Caldwell2023, shen2021} while CW-based frequency transfer has operated with up to 12\,dB of loss across turbulent terrestrial free-space channels~\cite{Gozzard2022}. With additional amplification of the CW laser, the latter technique could tolerate several more orders of magnitude of loss. These terrestrial experiments suggest that with sufficient aperture sizes, the link budgets for ground-to-space time transfer should not be prohibitive. A $10^{-18}$ comparison also requires the relativistic corrections to be modeled to the same level: in low Earth orbit the gravitational redshift changes by about $10^{-16}$ per meter of altitude and the second-order Doppler shift by about $10^{-16}$ per millimeter per second of velocity, so the satellite's altitude must be known to about 1\,cm and its velocity to about 10\,$\mu$m/s, at the limit of current precise orbit determination. 

As a satellite moves across the sky during an overpass, the changing line-of-sight pseudorange, or range-rate, will introduce Doppler shifts on the light. For a range-rate of 5\,km/s that may be seen in some LEO orbits, the fractional Doppler shift of the optical carrier will sweep between $\pm 2 \times 10^{-5}$ during a few-minute overpass. Terrestrial demonstrations have shown the ability to correct $10^{-7}$-level Doppler shifts for $10^{-18}$-level clock comparisons~\cite{bergeron2019, McSorley2025}, but the ability to perform clock comparisons in the presence of these much larger shifts remains an open problem. Ground-station infrastructure and link stabilization are discussed in the contributions by Schediwy and Dix-Matthews in Sec.~\ref{sec:proceedings}.

\subsubsection{Atom interferometers}
\label{sec:AI}

Atom interferometers exploit the wave-like nature of cold atoms to
measure phase shifts accumulated along distinct quantum trajectories.
The operating principle is analogous to an optical Mach--Zehnder
interferometer: a sequence of resonant laser pulses acts as beamsplitters
and mirrors for the atomic wave function~\cite{kasevich91corr, Cronin2009,
Buchmueller:2023interferometry}. In the standard $\pi/2$--$\pi$--$\pi/2$
pulse sequence, the first pulse places each atom into a coherent
superposition of two momentum states; the paths separate spatially over
an interrogation time $T$; a $\pi$-pulse redirects both arms; and a final
$\pi/2$-pulse recombines them, yielding an interference signal whose phase
encodes inertial effects such as acceleration and gravity. The
sensitivity scales with the square of
the interrogation time, making extended free fall a key figure of merit
for fundamental physics reach.

Large-scale instruments are motivated primarily by two classes of signal.
Ultralight dark matter (ULDM) with scalar couplings to Standard Model
fields produces oscillations in atomic transition frequencies, which can
be read out as a differential phase shift between two spatially separated
interferometers interrogated by a common laser~\cite{Arvanitaki2018}, in
a gradiometer configuration that simultaneously suppresses laser phase
noise~\cite{Badurina2020}. Vector couplings, for example to $B{-}L$,
instead produce differential accelerations between test masses of
different composition, which interferometers operating on two species
or isotopes measure directly~\cite{Graham:2015ifn}. Gravitational waves
act differently: a passing strain
modifies the light travel time between free-falling atom clouds and
produces a differential phase proportional to both the baseline $\Delta r$
and the effective momentum transfer~\cite{Graham2013}. Large momentum
transfer techniques and quantum-enhanced readout further extend the
achievable sensitivity~\cite{wilkason2022atom, hosten2016measurement}.

\begin{figure}[h!]
\begin{center}
\includegraphics[width=0.45\textwidth]{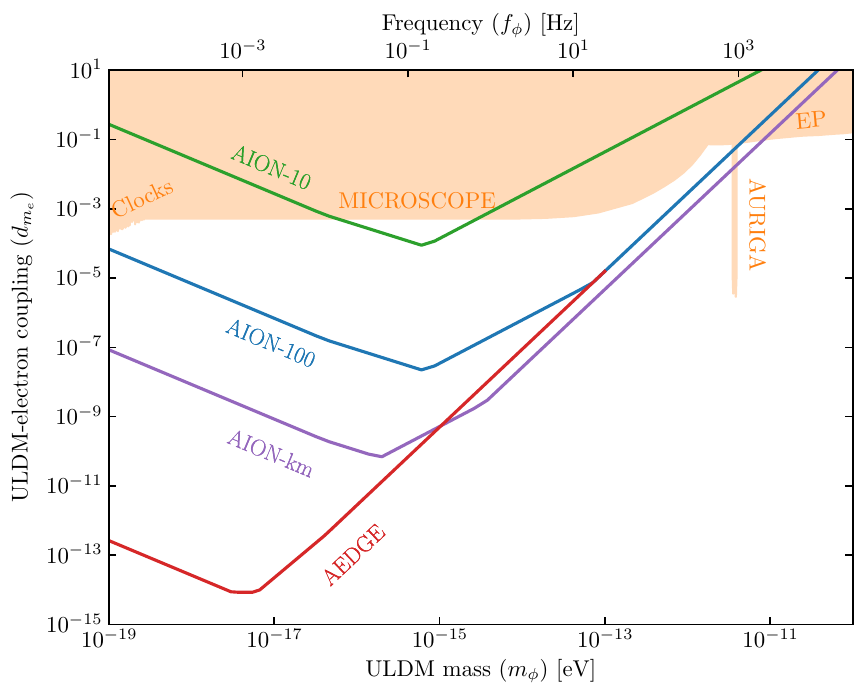}
\hfill
\includegraphics[width=0.53\textwidth]{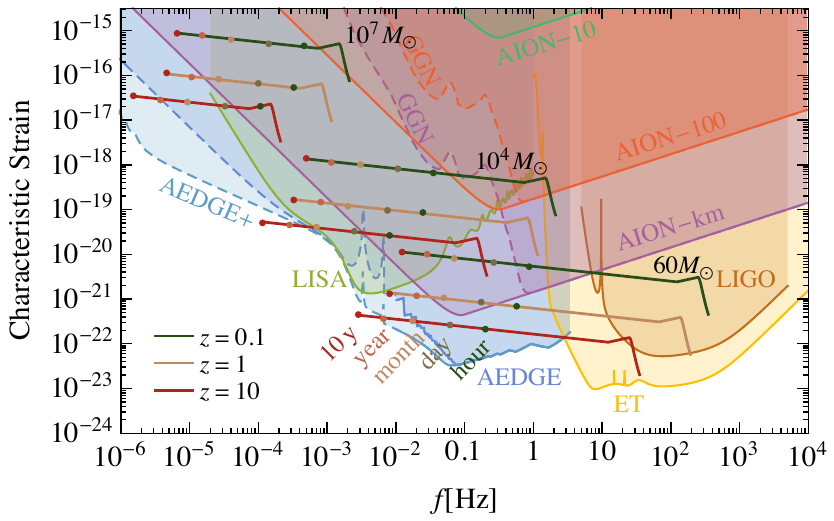}
\caption{\label{fig:summary-AI}
Projected sensitivities of large-scale atom interferometers to ultralight dark matter
(left, adapted from~\cite{Badurina2020}) and gravitational waves (right,
from~\cite{Badurina:2021rgt}). Terrestrial pathfinders now under
construction (AION, MAGIS, and others) will provide the first meaningful
reach in both channels, with sensitivity improving progressively toward
km-scale detectors. The ultimate sensitivity becomes accessible once the
technology is brought into space, as illustrated here by one
representative mission concept, AEDGE.
}
\end{center}
\end{figure}

The terrestrial landscape of large-scale prototypes has grown
substantially~\cite{Buchmueller:2023interferometry}. Instruments at the
10--100\,m scale are operating or under construction at
multiple sites: AION-10 at Oxford~\cite{Badurina2020}, MAGIS-100 at
Fermilab~\cite{Abe2021}, MIGA in France~\cite{Canuel2018},
VLBAI at Hannover~\cite{schlippert2020matter}, and ZAIGA in
China~\cite{Zhan:2019quq}. The differential architecture on which these
instruments rely has now been demonstrated, with a prototype strontium
clock interferometer showing the common-mode rejection of laser
frequency noise that long-baseline operation
requires~\cite{Baynham2026}. The next step in scale is the proposed AICE
facility at CERN~\cite{AICE:2026}, which would take this architecture to
the 100\,m baseline. Taken together, these efforts mark a
transition from laboratory-scale sensors to distributed observatories,
establishing the foundations for km-scale detectors that would open the
decihertz band at $10^{-2}$--$1$\,Hz, as discussed above,
while delivering competitive sensitivity to ULDM~\cite{Badurina:2021rgt}.
A community roadmap and proto-collaboration have been established with the
goal of commissioning at least one km-scale instrument by the
mid-2030s~\cite{TVLBAI:2023, Alonso:2022oot}.

Mission concepts such as AEDGE~\cite{El-Neaj2020}, proposed to ESA in
response to the Voyage 2050 call, would extend this program to space,
probing gravitational waves across the decihertz band and ULDM in the mass range $10^{-21}$--$10^{-15}$\,eV
with sensitivities well beyond those of any terrestrial
instrument~\cite{Badurina:2021rgt}. A conceptual feasibility study for a
100-m-scale precursor at CERN has been
completed~\cite{Arduini:2023wce}, and a technical proposal has since been
submitted for the Atom Interferometer CERN Experiment
(AICE)~\cite{AICE:2026}, a vertical instrument of order $100$\,m to be
installed in the PX46 access shaft to the LHC. AICE would operate
ultracold $^{87}$Sr on the $698$\,nm single-photon transition in a
multi-source gradiometer configuration, targeting bosonic ultralight dark
matter in a mass range inaccessible to other experiments and, as a
secondary goal, gravitational waves at $0.03$--$3$\,Hz, with later
operation on $^{88}$Sr and $^{171}$Yb extending the reach to axion-like
particles, $B{-}L$ vector couplings and tests of the equivalence
principle. Endorsed by the TVLBAI proto-collaboration and sited so that
the preparatory works fall within the third long shutdown of the LHC,
it is the concrete near-term
milestone on the path to km-scale and space-based detectors.
As illustrated in Fig.~\ref{fig:summary-AI},
the terrestrial pathfinder program already under construction will
deliver meaningful sensitivity to both signal classes, while the ultimate
reach in both ultralight dark matter and gravitational waves becomes
accessible once the technology is brought into space. The community
roadmap for this program is described in the contribution by
Buchmueller; space-based pathfinders and mission concepts in the
contributions by Gaaloul and Williams; atom sources and interferometry
techniques in the contributions by Marchant, Tien, Glick, and Periwal;
and monitoring of the upper atmosphere in a second contribution by
Marchant, all in Sec.~\ref{sec:proceedings}.

\subsection{Outlook}

The science cases make convergent demands on the sensors. For clocks,
a single trapped-ion clock with $10^{-15}/\sqrt{\tau}$ instability and
$10^{-18}$ inaccuracy, paired with an optical time-frequency link of
comparable performance, meets the equivalence-principle, microhertz
gravitational-wave, VLBI, and non-common-view comparison cases
(Fig.~\ref{fig:summary-clocks}); every dark-matter search gains from
improved stability; and only decihertz gravitational-wave
detection with clocks demands more. Such a clock can be built from
current technology in about 20\,L and 100\,W. For atom interferometers,
the equivalence-principle and dark-matter cases call for the longest
free-fall times, and the gravitational-wave and gradiometer cases for
baselines of order $10^{4}$\,km (Fig.~\ref{fig:summary-AI}).

Pathfinder missions are the near-term priority. For clocks, a complete
single-ion optical clock in medium Earth orbit, in a higher geocentric
orbit, or on the lunar surface, compared with ground clocks through an
optical link, would improve the gravitational-redshift test by two to
three orders of magnitude, search for ultralight dark matter with
particle masses from $4\times10^{-23}$ to $4\times10^{-12}$\,eV and for
transient dark-matter objects, and qualify the cavity-stabilized laser,
frequency comb, ion trap, and link terminal that heliocentric and
planetary missions will need. For atom interferometers, the Cold Atom
Lab and the CARIOQA accelerometer pathfinder play this role, STE-QUEST
is the near-term science mission, and the terrestrial program, from
today's 10\,m instruments through the proposed AICE facility at
CERN~\cite{AICE:2026} to a km-scale detector by the mid-2030s,
leads toward AEDGE- or MAGIS-Space-class detectors. ACES, now operating on the ISS, does the
same for microwave clocks and links.

Several developments must proceed alongside the sensors. Optical
time-frequency transfer at $10^{-18}$ has been demonstrated over
terrestrial links but not yet at the Doppler shifts of a
ground-to-space link, and every clock-based case except onboard
clock--clock or clock--cavity comparisons depends on it. Photonic integration of lasers,
modulators, and traps is a promising route to the size, weight, power, and
robustness that flight requires. The theory inputs that set the
targets, from the dark-matter density in the inner solar system to the
signatures of transient fields, also deserve continued attention.

Finally, several agency planning processes are open now. In the United
States, the most recent NASA Biological and Physical Sciences Division decadal survey proposes a
research campaign, Probing the Fabric of Spacetime, built around clocks
and atom interferometers in orbit~\cite{BPSdecadal2023}; NASA's
Astrophysics Division is currently identifying its next strategic missions
through the ASTRA initiative and its Ad ASTRA community
workshop~\cite{AdASTRA2026}; and, outside NASA, the 2023 report of the
Particle Physics Project Prioritization Panel (P5), which advises DOE
and NSF, recommends a portfolio of small, agile dark-matter experiments
using new technologies~\cite{P5report2023}. In Europe, the community roadmap
submitted to ESA for Voyage 2050~\cite{Alonso:2022oot} predates advances in optical-clock development and is due for an update.
Coordinated input to all of these, ahead of the next decadal surveys,
would present NASA, DOE, and ESA with one unified program.

\section{Proceedings}
\label{sec:proceedings}

The remaining sections of these proceedings present one-page contributions
from the speakers and poster presenters at the workshop, each
summarizing the talk or poster on which it is based. Talks appear
first and posters second, in the order in which they were presented.
Each contribution is written by its presenter and represents that
author's own perspective, in contrast to the preceding summary, which
was prepared collectively. The author list of these proceedings
comprises the workshop participants; co-authors of a contribution who
did not attend the workshop are credited in the heading of that
contribution. Together with the preceding summary, they document the science cases, mission concepts, and
technology developments discussed at the Aspen Center for Physics workshop on
``Fundamental physics in space with quantum sensors.''

\subsection{Discovering new physics with quantum technologies in space \\
            Marianna Safronova}

A broad class of well-motivated dark-matter candidates consists of
ultralight bosonic fields with masses far below one electronvolt. In
this regime, the number of dark-matter quanta within a de~Broglie
volume is so large that the field is best described as a coherent
classical oscillation. For example, a dark-matter particle of mass
$10^{-16}$~eV oscillates at approximately $24$~mHz with a coherence
time of months, and its de~Broglie wavelength is comparable to the
size of the Solar System, so a network of clocks distributed across
multiple spacecraft would see a single coherent signal.

If the dark-matter field couples weakly to the electromagnetic
interaction or to the mass of the electron, it produces small,
periodic oscillations in the fundamental constants of nature, such as  the fine-structure constant $\alpha$ and the
electron-to-proton mass ratio. Different atomic transitions respond
differently to these variations, so the time-dependent ratio of two
clock frequencies allows to extract the dark matter signal.  By comparing clocks based on different elements, one can also
search for transient signals from dark-matter clumps such as
topological defects~\cite{Derevianko2014}. The Global Positioning
System constellation, treated as a $50{,}000$-km clock array
gravitationally bound to Earth, has already been used in this way to
search for domain-wall dark matter, setting limits that exceed
astrophysical bounds by several orders of magnitude on certain
couplings~\cite{Roberts2017}.

The local density of dark matter in the Solar System is not directly
known. Estimates based on galactic averages give roughly
$0.4\,\mathrm{GeV/cm^{3}}$, but theoretical work has shown that overdensities may exist
bound to the Sun or to giant planets that exceed the average by many
orders of magnitude~\cite{Budker2023}. A satellite mission
carrying two atomic clocks of different element compositions,
delivered to about $0.1$~astronomical units from the Sun, would (i)
directly probe the existence of a solar dark-matter halo, (ii) explore
parameter space for the relaxion, a theoretically motivated
ultralight scalar that simultaneously addresses the Higgs hierarchy
problem and supplies a viable dark-matter candidate, and (iii)
detect any spatial variation of the fundamental constants correlated
with the gravitational potential of the Sun~\cite{Tsai2023,Budker2025}.

Scalar dark-matter models with quadratic coupling lead to screening that the field is 
suppressed near a massive body, so the dark-matter amplitude
available at the Earth's surface is much smaller than the cosmic
value. Recovering the unscreened amplitude requires getting  away from the Earth surface beyond LEO orbit~\cite{Arakawa2024,Banerjee2023}.

The FOCOS (Fundamental physics with a state-of-the-art Optical Clock
in Space) mission concept~\cite{FOCOS2022} proposes flying a
strontium optical lattice clock on a highly elliptical Earth orbit
between altitudes of $5{,}000$ and $22{,}800$~km, optically linked to
clocks on the ground. The primary scientific goal is a test of the
gravitational redshift,
\begin{equation}
\frac{\Delta\nu}{\nu} \;=\; (1+\alpha)\,\frac{\Delta U}{c^{2}},
\end{equation}
in which any deviation of the parameter $\alpha$ from zero would
indicate a violation of a foundational prediction of GR. FOCOS targets
a sensitivity on $\alpha$ that is roughly $30{,}000$ times better than
the best current limit. Additional FOCOS objectives include
other tests of relativity, enhanced dark-matter searches that take
advantage of the varying gravitational potential along the orbit,
improved limits on the slow drift of fundamental constants over time,
and the establishment of a high-accuracy international time and
geodetic reference~\cite{FOCOS2022}.

Existing and planned gravitational-wave observatories cover several
distinct frequency bands and atomic clocks in space can fill the gaps to 
observe the GW between ground based laser interferometers and LISA as well as between LISA and pulsar timing arrays. 
 The CHRONOS concept~\cite{Kolkowitz2016} places identical
optical lattice clocks on two drag-free satellites that share an
ultrastable laser link. A passing gravitational wave imprints a
differential frequency shift on the link, which the clocks read out
with sufficient precision to detect compact-binary inspirals. A
particularly compelling application is the localization of
binary-neutron-star inspirals on the sky days to weeks before merger,
allowing electromagnetic observatories to point at the source in
advance of the resulting kilonova explosion. Another complementary
concept uses asteroids in the inner Solar System, which are naturally
drag-free, gravitationally quiet test masses, linked by
inter-asteroid laser ranging to reach gravitational-wave frequencies
near $10\,\mu\mathrm{Hz}$~\cite{Fedderke2022}.

Multimessenger astronomy already combines photons, neutrinos, cosmic
rays, and gravitational waves from a common astrophysical source.
Quantum sensors can add a new channel: bursts of ultralight bosonic
fields predicted by some extensions of the Standard Model. Compact
binary mergers, core-collapse supernovae, and the collapse of bosonic
field configurations (``bosenovae'') are all candidate sources. A
network of clocks and magnetometers can register such a burst as a
correlated transient arriving slightly after, but in coincidence with,
the corresponding electromagnetic or gravitational-wave signal,
opening a new window onto physics beyond the Standard
Model~\cite{Arakawa2026}.

\subsection{CHRONOS: A Quantum Clock Space Telescope for Gravitational Astrophysics \\
            Peter Kurczynski}

CHRONOS \footnote{Chronometric High-precision Relativistic Observatory of Networked Optical-clocks in Space} is a notional spacecraft mission concept consisting of a quantum-clock telescope to detect gravitational waves. Although gravitational waves were first detected in 2015, and were the subject of Nobel prizes in 2017, the gravitational-wave sky is still largely unexplored and detecting them is extremely difficult. The ground-based Laser Interferometer Gravitational Wave Observatory (LIGO) first detected high-frequency gravitational waves \cite{Abbott2016}, and the planned Laser Interferometer Space Antenna (LISA) will probe low-frequency gravitational waves from space \cite{Babak2021}. 

CHRONOS would enable detecting gravitational waves in the mid-frequency band, referred as  the \textit{decihertz} (dHz) band, ($\sim$ 0.1 - 10 Hz). Sources of emission in these frequencies are believed to include merging compact-object binaries, especially neutron stars, in the months leading up to their merger and subsequent explosion. This frequency band is also ideally suited for precise on-sky localization of gravitational-wave sources. Therefore, CHRONOS would provide an early warning for explosive transients such as kilonovas and gamma-ray bursts, giving astronomers precise knowledge of where and when these astrophysically interesting transients will occur. 

Unlike current and planned gravitational-wave detectors, which use interferometry to measure minuscule expansions and contractions of space, this concept will use frequency shifts between widely separated clocks to measure minuscule changes in time. Although the requirements for optical clock accuracy are several orders of magnitude beyond the current state of the art, optical clock technology is rapidly improving. Continued progress in the coming decade may enable this mission concept.

CHRONOS would consist of two widely separated, drag-free spacecraft in heliocentric orbit. They would be connected by a highly stable laser optical link, enabled by an onboard laser frequency comb. Each spacecraft would be equipped with an optical clock to provide high precision timing reference.

Back of the envelope calculation suggests that (strain) sensitivity to gravitational waves of such a two-spacecraft architecture varies linearly with spacecraft separation and clock accuracy (fractional frequency shift). Spacecraft separation of $\sim 10^6$ km and clock accuracy of $\sim 10^{-21}$ combine to yield strain sensitivity of $\sim 10^{-21}$, which is suitable for gravitational wave detection. Likewise, preliminary computation of the gravitational wave transfer function suggests that these mission parameters can deliver plausible sensitivity across the mid-band frequency spectrum. 

Optical clocks have been proposed by atomic physicists as gravitational wave detectors \cite{Kolkowitz2016}. There is now a need to establish their feasibility for spaceflight by investigating them in more detail with experts in gravitational-wave detection, space-based science mission architecture and spaceflight qualification of optical clock technologies. 

The methodology of Concept Maturity Levels (CMLs) is becoming a best-practice in the early development of science spacecraft mission design and architecture \cite{Wessen2013,Ziemer2018}. This methodology advances concurrently science objectives, mission design, technical risk, project organization, and cost. CMLs are being used for NASA Astrophysics' newest flagship mission concept, the Habitable Worlds Observatory \cite{Lee2026}. 

CMLs are a graduated set of levels intended to gauge a mission concept's early development, analogous to Technology Readiness Levels that are now an engineering standard. Each CML is characterized by specific concerns and criteria. CMLs systematize the mission concept development process and help concept development teams streamline their activities by focusing on the right conceptual problems at the right time. 

The mission concept development process begins with CML 1. At this stage, the mission may be understood only at the level of a ``cocktail napkin" idea or concept. Progress begins with establishing overall goals and objectives for the mission. These will be refined as the concept matures. In this context, goals are generally broad, qualitative in nature, and unlikely to be accomplished by a single mission. They place the particular mission in context. Objectives are more specific, and they will ultimately be honed into precise statements. Objectives lead to Level 1 requirements of a mature project. NASA defines success as when a mission has accomplished its objectives. Therefore objectives must be framed in a way that can be evaluated unambiguously, and ideally quantitatively.

The scientific method provides a useful structure for framing objectives. In this method, questions are formulated that lead to hypotheses, which generate predictions that can be tested by observations. Science advances through confirming or refuting hypotheses through experimentation and (more typically in astrophysics) observation.

Hypothesis driven science beats measurement capability driven science. Objectives that are motivated by testing a specific hypothesis will generally yield stronger mission concepts than objectives that seek purely to improve upon existing measurement capability. Objectives that are purely measurement-capability driven will tend to be harder to justify to stakeholders. Such objectives tend to be harder to defend as mission complexity increases and inevitable tradeoffs must be envisioned between technical capability on the one hand and technology readiness, cost, schedule and risk on the other. Thus a central challenge of conceiving missions that seek to extend observational capability into new domains is to frame compelling hypotheses that can be tested with feasible observations.  

\subsection{Asteroids for the detection of fundamental physics in the $\mu$Hz band \\
            Michael A. Fedderke}

The advent of gravitational-wave (GW) detection~\cite{Abbott2016} has opened a new window on the universe, but we have only just begun to peel back the curtain on the discoveries that this new probe will enable. 
It is imperative to observe as much of the GW frequency spectrum as possible. 
Indeed, a broad effort is ongoing on this front; however, a gap exists in proposed GW detection techniques in the $\sim 10^{-7}$--$10^{-4} \, \text{Hz}$ frequency band, below LISA~\cite{Colpi2024} but above that of pulsar timing arrays~\cite{Agazie2023,Reardon2023,Xu2023,Antoniadis2023}.  
There is no dearth of interesting astrophysical and cosmological GW sources in this band~\cite{Sesana2019}: observations here will certainly hold interesting lessons for those fields, and possibly even for fundamental particle physics.  
As severe low-frequency noise limitations inhibit accessibility of this long wavelength/low-frequency GW band from the ground, a transformative science opportunity exists for space-based missions, well aligned with Key Scientific Questions Theme 3 and Key Science Question 11 in the 2023 BPS Decadal Survey~\cite{BPSdecadal2023}.

My collaborators and I have proposed the use of asteroids as test/proof masses for a GW detector~\cite{Fedderke2022}. 
The key idea is to instrument certain carefully selected asteroids with a human-engineered timing/link system to permit monitoring of inter-asteroid separation distances.
Such instrumentation would most likely take the form of an asteroid orbiter that references itself to a network of retroreflectors deployed to the surface of a local asteroid, and which communicates across a long baseline to a similar spacecraft located in orbit around another asteroid, permitting inter-asteroid ranging; other configurations involving more than two asteroids are also possible.
Because a GW causes periodic fluctuations in the proper distance between test masses, such a system constitutes a single-baseline GW detector. 
In particular, we demonstrated that a mission of this type may be able to observe the exceedingly challenging {$0.3$--$30\,\mu\text{Hz}$} GW frequency band.
Such a detector would also be well positioned to explore other aspects of fundamental physics, such as dark matter candidates that exert subtle forces on ordinary matter~\cite{Baum2022,Fedderke2023}.

It is necessary to go to space to observe in this frequency band because the backgrounds on Earth are far too large (e.g., Newtonian noise from Earth's vibrations coupling gravitationally to the detector).
Even in space, however, engineering sufficiently stable proof masses at these low frequencies would require environmental isolation far beyond the level achieved even by the LISA Pathfinder mission~\cite{Armano2018}.
We showed, on the other hand, that appropriate asteroids are sufficiently insensitive to known and characterized space-environmental perturbations as to permit them to serve as natural, stable test masses that can be exploited with the requisite timing/link system~\cite{Fedderke2022}.  

The major scientific unknowns associated with realizing the science case for a mission of this class lie in two categories: (1) evaluating the tradespace between the number of asteroids to be employed as test masses and the technologies required and available to exploit them, and (2) further characterization of asteroids as test masses.

Regarding point (1), possible mission architectures include: round-trip timing across a single baseline (two asteroids) referenced to a high-precision space-qualified atomic clock; relative round-trip timing across two or more baselines (three or more asteroids) without the need for an atomic clock; or relative interferometric measurements across two or more baselines, again without an atomic clock.
Trade-offs exist between these cases in terms of mission complexity and technological sophistication.
Space-qualified atomic clocks in the $\delta \nu / \nu \sim 10^{-18}$ frequency-stability class are required for the single-baseline architecture; such stability has been demonstrated on the ground~(e.g., Ref.~\cite{Bothwell2019}), but their space qualification and autonomous operation is an area of intense ongoing R\&D work (and Ref.~\cite{BPSdecadal2023} calls out such clocks on orbit as a transformative technology).
On the other hand, link-system architectures for a single baseline could in principle be as simple as a pulsed laser beacon emitted from one asteroid orbiter, with active transponder feedback from the orbiter at the distant asteroid.
Multiple baselines would alleviate the requirement for a space-based atomic clock, but may be more technically complex to engineer in terms of the link systems and mission profile, especially for interferometric measurement setups (although, LISA-like laser systems are mature technology).

Regarding point (2), a major open question for this proposal's viability is that direct, in-situ measurements of the seismic and lower-frequency / long-term surface stability of asteroids have not to-date been made.
Further study of this question, as well as characterization of viable asteroid targets in terms of their orbits, geology, rotational motion, etc. is crucial for advancing this mission.
This would be synergistic with explorations of interest both to planetary defence applications (cf., the recent DART mission~\cite{Daly2023a,Daly2023b}) and to communities engaged in understanding the origin and evolution of planetary systems, as well as the particular origins of our solar system itself.

A mission of this class is clearly technologically sophisticated, but many individual aspects of the technologies it would need to leverage have been demonstrated in other settings.
For instance, ground-based atomic clocks of the requisite class have been demonstrated, but space-qualified clock technologies are not yet at the level required.
NASA, JAXA, and ESA have all previously orbited and/or made soft landings on either asteroids or comets, in some cases even achieving sample return.
Acquisition of laser pulses retroreflected from Moon landers and their use in orbit determination, a technology necessary for an orbiter to reference itself to a local asteroid, has also recently been demonstrated~\cite{Cascioli2025}.
And there is a large body of work on space-based laser and radio link systems, although none currently optimized for a mission in this class.
Importantly, as they mature, many of the technologies required for this mission could be tested on orbit, independently and in parallel, in ways motivated by their own science cases.

\subsection{Fundamental Physics Across Scales: From Astrophysical Environments to Quantum Sensors \\
            Volodymyr Takhistov}

\begin{figure}[!h]
  \centering
  \includegraphics[width=0.70\textwidth]{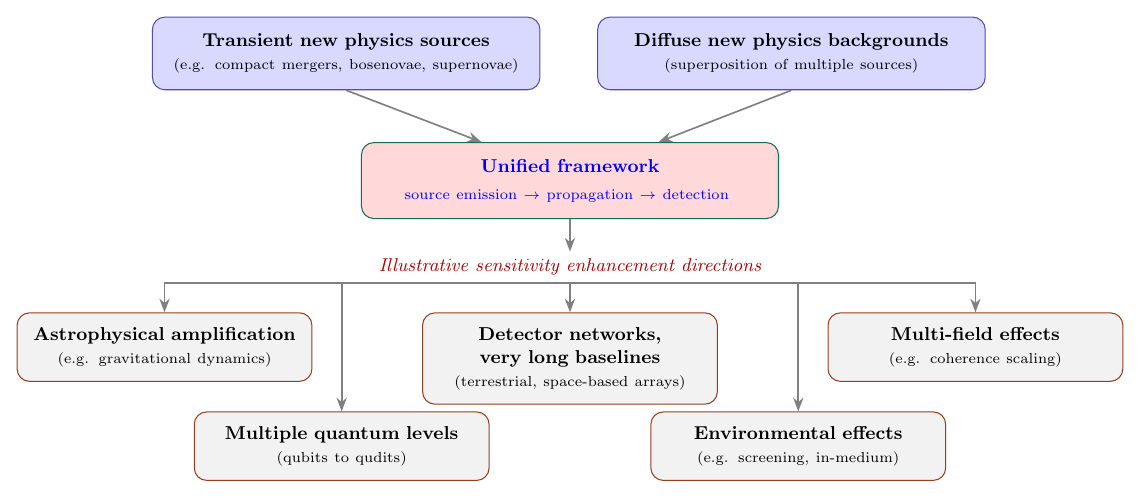}
  \caption{Schematic illustration of a unified framework connecting astrophysical source emission, propagation, and detection of novel quantum fields, together with illustrative directions for sensitivity enhancements.}
  \label{fig:Lastname-figname}
\end{figure}

Searches for fundamental physics beyond the Standard Model benefit from leveraging quantum field manifestation in qualitatively distinct physical regimes such as individual particle-like quanta, classical oscillatory backgrounds or transient configurations as well as static forces. Each realization leads to distinct observables and requires corresponding experimental strategies. Connecting insights across traditionally separate domains, such as astrophysics and precision quantum sensing, establishes a unified pathway for probing fundamental interactions.

Extreme astrophysical systems act as natural amplifiers of weak interactions, providing sensitivity to microscopic physics complementary to laboratory experiments. For example, supermassive black hole binary dynamics constrain ultralight bosons near masses of $\sim 10^{-21}$~eV through superradiant effects~\cite{Ding:2025nxe}, while diffractive gravitational wave lensing probes small scale dark matter structure around compact objects~\cite{GilChoi:2023ahp}. More broadly, the emergence of multimessenger astronomy from the coincidence observation of supernova neutrinos~\cite{Kamiokande-II:1987idp,Bionta:1987qt} and photons from SN1987A and the joint gravitational wave and electromagnetic detection of neutron star merger GW170817~\cite{LIGOScientific:2017ync} demonstrated that violent astrophysical events provide powerful testbeds for fundamental physics through complementary observables.

Leveraging this paradigm, we have developed a unified framework connecting astrophysical source emission, propagation and detection of new light fields within a single pipeline~\cite{Eby2021}, exemplified for axion star bosenovae. Populations of sources can generate diffuse backgrounds composed of novel fields~\cite{Eby:2024mhd}. Fields produced in extreme environments such as  compact object mergers, supernovae and bosenova explosions  propagate as wave packets whose spectral structure can encode underlying fundamental   interactions, production mechanisms and propagation history.
Notably, detection sensitivity is not determined solely by experimental performance, but includes the interplay with source emission and propagation. This extends the concept of multimessenger astronomy beyond traditional channels and searches.

Quantum sensors enable sensitive probes of such weak interactions and this framework naturally extends to distributed quantum sensor networks~\cite{Dailey2020}, which act as multimessenger receivers sensitive to both transient and continuous field configurations.  Coincidence analyses across gravitational wave, electromagnetic, and neutrino channels enable probing a broad class of new interactions~\cite{Arakawa2023,Arakawa2024,Arakawa2026}. Space-based platforms can further enhance sensitivity, such as through extreme, distinct backgrounds, systematics and   possible environmental effects (e.g. screening)~\cite{Stadnik2021,Derevianko:2021wgw,Arakawa2024,Arakawa2026,delCastillo:2025rbr}. A wide range of quantum sensing platforms, including precision clocks, can exploit such signals. Enhancements and modified physical effects can also arise from multi-field effects as exemplified by axions~\cite{deGiorgi:2025ldc} as well as from leveraging  multiple quantum levels in sensing systems beyond two-state qubits~\cite{Ma:2025orp}. Strong complementarity of terrestrial and space-based searches enabled by quantum sensing technologies is well illustrated by transition edge sensors, expected to be deployed in underground laboratories for dark matter searches~\cite{Chen:2025cvl,Schwemmbauer:2025evp}, sub-orbital missions and future space-based observatories~\cite{Matsumura:2013aja}, bridging distinct measurements across different physical targets.

These developments indicate that searches for new fundamental physics are inherently multi-dimensional and progress will increasingly rely on integrating information across environments, field realizations and sensing platforms, establishing a coherent framework that connects distinct phenomena across measurement scales.

\subsection{Quantum Sensors for Monitoring of the Upper Atmosphere \\
            Anna L. Marchant}

Quantum sensors based on atom interferometry have found applications in a range of fundamental physics settings, from tests of the equivalence principle \cite{Overstreet2018} to determination of the fine structure constant \cite{Morel2020}. Future experiments aim to exploit such devices to detect gravitational waves and the effects of dark matter \cite{El-Neaj2020, Badurina2020, Abe2021} using extended baselines to improve sensitivity. Similarly, the enhanced precision offered by these quantum inertial sensors is also useful for Earth observation applications, using atoms as drift-free accelerometers.

Understanding the dynamics of the Earth's atmosphere is key to weather and climate modelling. The thermosphere, a layer of the Earth's atmosphere stretching from approximately 100~km to 600~km, is of particular importance as its dynamics are driven by solar extreme ultraviolet emissions and the interaction of the solar wind with Earth’s magnetosphere. This results in temporal and spatial density variations of the thermosphere which in turn impacts energy redistribution within the atmosphere as a whole. Collecting data on these variations is of great interest for future climate models. 

The effect of the upper atmosphere on the lifetime of low Earth orbiting satellites is another key area of interest due to the drag experienced in flight. With the increase in the number of satellites being launched into low Earth orbits comes the need to better understand, and better model, the upper atmosphere. Variations in density lead to uncertainty in satellite orbit predictions, potentially increasing the risk of collisions with other objects. Measurements of the atmospheric density, based on atmospheric drag, could therefore also be invaluable for improved orbit predictions. 

The Cold Atom Interferometric Thermosphere Drag Measurement (CAITDM) project has developed a breadboard model of a cold atom accelerometer, designed to perform atmospheric drag measurements on-board a 16U CubeSat. Targeting data collection in a previously unexplored parameter regime (in terms of altitude and solar activity levels) \cite{Siemes2022}, the project also acts as a technology demonstrator for free-flying cold atom space systems.

\begin{figure}[h!]
\begin{center}
\includegraphics[width=0.8\textwidth]{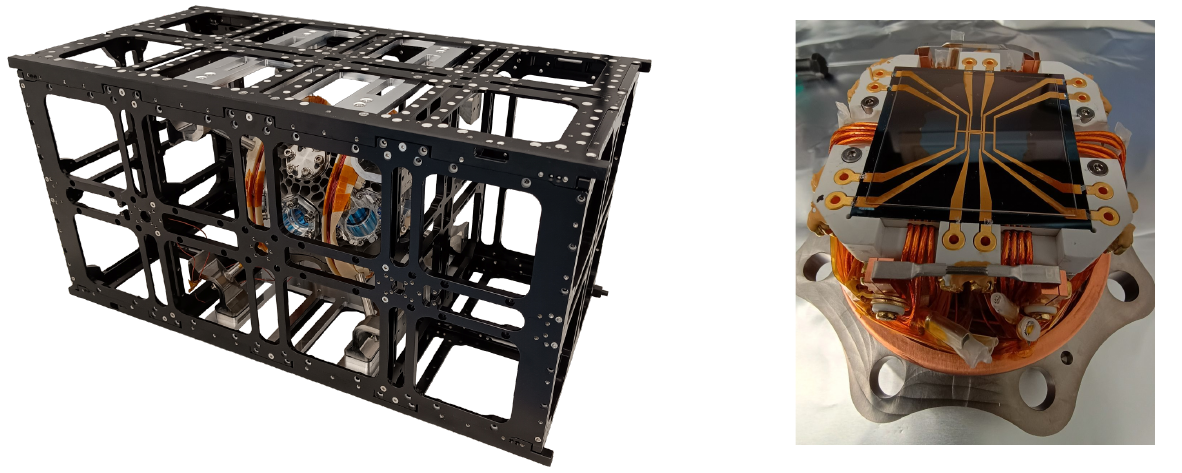}
\caption{\label{fig:Marchant-CAITDM_fig}
    The CAITDM system. Left: Vacuum chamber mounted inside the 16U CubeSat frame. Right: Atom chip, developed at RAL Space, mounted on the lower vacuum chamber flange that incorporates all required electrical feedthroughs.  
}
\end{center}
\end{figure}

In order to reduce Size, Weight and Power (SWaP), a number of innovative design elements have been incorporated into the system. The CAITDM science package is comprised of a 3D printed vacuum chamber, shown in Fig.~\ref{fig:Marchant-CAITDM_fig}, reducing the mass by around 50\% compared to a traditional chamber machined from solid metal. This houses an atom chip device to generate magnetic fields close to the atoms. Compact laser delivery optics are attached to the chamber, along with both passive and active magnetic shielding components. In contrast to bulky conventional magnetic coils, quantisation and compensation fields are generated by intricately designed printed circuit boards \cite{Davis2024}. The quantum sensor itself is based upon rubidium atoms, requiring frequency stabilised laser light at 780~nm. This light is generated by frequency-doubling two 1560~nm lasers, driven by custom electronics. The first is used as a reference, stabilised using modulation transfer spectroscopy, and the second is tuned via an offset lock to give the laser cooling, interferometry and imaging light.

The advantage of quantum sensors for Earth observation lies not only in the exquisite control achievable in atomic system but also the reproducibility afforded by the use of atoms. Unlike classical sensor which suffer short-terms drifts, requiring frequent calibration, atoms act as inertial test masses as well as absolute frequency references, resulting in high accuracy. Not only does this make them ideal for atmospheric drag measurements but the same accelerometers can be applied to gravity field and gravity gradient mapping of the Earth \cite{Trimeche2019}.

\subsection{Fundamental physics in the lab, in an elevator and in space \\
            Changhao Xu, Wei Ji, and Dmitry Budker}

Levitated ferromagnetic systems have recently emerged as a powerful platform for precision sensing, combining concepts from quantum metrology, spin physics, and macroscopic mechanics. In these systems, a ferromagnetic particle behaves as a collective “macrospin,” whose precession in an external magnetic field can be monitored with high precision\,\cite{kimball2016precessing}. Unlike ensembles of independent spins, ferromagnets enable a collective enhancement of signal strength, opening a pathway toward sensitivities that can surpass the standard quantum limit.

Recent experimental advances have demonstrated levitated ferromagnetic sensors operating in ambient and controlled environments, achieving magnetic field sensitivities at several tens of fT$/\sqrt{\mathrm{Hz}}$\,\cite{ji2025levitated}. These systems rely on stable trapping and precision optical readout for magnet's motion and orientation. In particular, feedback-controlled magnetic traps enable robust operation even under free-falling conditions, which is essential for realistic deployment beyond laboratory environments.

A key enabling technology is a tunable, feedback-stabilized magnetic trap. A recently demonstrated master PID magnetic trap combined optical position detection, real-time feedback, and a master control coil to levitate a ferromagnetic particle in the Einstein Elevator. The system remained stable through release shocks and resolved the magnet's motion in both reduced-force and pure-free-fall configurations~\cite{Xu2026}. This capability is an important step toward free-precession measurements and space-based ferromagnetic magnetometry.

The relevance of levitated ferromagnetic sensors for space applications lies in their potential for long coherence times and reduced environmental noise. Microgravity conditions naturally suppress mechanical coupling, enabling improved isolation and extended interrogation times. These features make levitated systems particularly attractive for precision measurements of weak effects, including tests of general relativity and searches for new physics.

One promising direction is the concept of a “gravity probe spin,” where the precession of a ferromagnetic macrospin is used to measure relativistic frame-dragging (Lense–Thirring effect) or geodetic precession\,\cite{fadeev2021gravity}. Compared to traditional gyroscopes, ferromagnetic sensors offer the possibility of enhanced sensitivity through collective spin dynamics and reduced technical noise. Sensitivity projections indicate that such systems could approach regimes relevant for detecting relativistic spin precession in compact experimental platforms.

In addition to gravitational physics, levitated ferromagnets can search for ultralight axionlike and dark-photon dark matter through oscillating physical or effective magnetic fields~\cite{Kalia2024}. Cryogenic levitation has demonstrated an energy resolution well below $\hbar$ and promising projected sensitivity to axionlike dark matter~\cite{Ahrens2025}, while double-resonance operation has enabled a broadband direct search for axionlike dark matter from 40 to 3000~Hz~\cite{Peng2026}. These measurements complement existing atomic and optical approaches.

Overall, levitated ferromagnetic sensors represent a versatile and scalable approach to precision metrology in space. Ongoing developments in trapping, control, and readout are rapidly improving their performance and robustness. With continued progress, these systems could play a significant role in future space missions targeting fundamental physics, including precision tests of gravity and searches for physics beyond the Standard Model.

\subsection{NASA's Cold Atom Lab on the ISS as a Pathfinder for Addressing Key Science Questions \\
            Jason Williams}

NASA’s Cold Atom Lab (CAL) on the International Space Station (ISS) is a multi-user facility enabling ultracold-atom experiments in the persistent microgravity environment of Earth's orbit. CAL is remotely commanded from the ground using a PXI-based control system, only requiring astronaut support for occasional onboard hardware replacements. The instrument draws \SI{565}{W} at \SI{28}{VDC}, fits within \SI{0.4}{\meter^3}, and has a mass of \SI{233}{\kilo\gram}, delivering a low-SWaP, high-stability quantum laboratory in orbit~\cite{Oudrhiri2024}. Operating since 2018, CAL has matured core quantum technologies for space, including atomic-referenced laser systems, ruggedized optomechanics, compact physics packages, and low-noise control electronics and current drivers, while executing more than 100{,}000 experimental runs for five research teams. These capabilities have enabled space-based pathfinder studies in atom interferometry (AI), dual-species quantum gas mixtures, quantum gas engineering, and ultracold atoms in unique (e.g., bubble) topologies. 

\begin{wrapfigure}{r}{0.5\textwidth}
  \centering
  \includegraphics[width=0.5\textwidth]{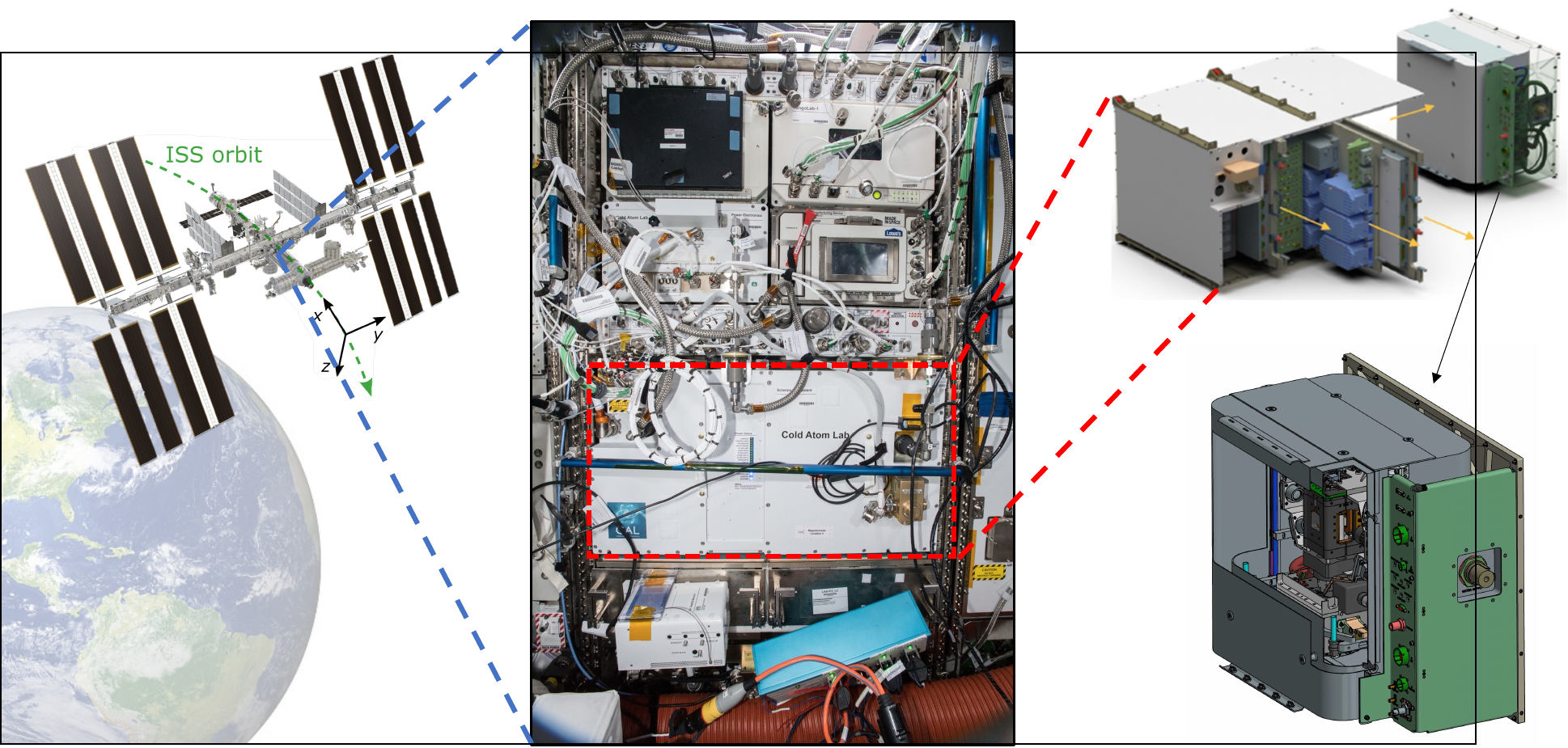}
  \caption{CAL occupies one quad locker (red dashed) and one single locker within an EXPRESS rack in the ISS Destiny module (center). The quad locker (upper right) includes the majority of the lasers, optics, current drivers, and control electronics. It also houses the science module (lower right) that consists of the ultra-high vacuum physics package, optomechanical bench, and dual-layer magnetic shields. CAL has operated robustly and reproducibly for over eight years, creating quantum gases of rubidium and potassium and supporting multiple research campaigns for the flight principal investigators.}
  \label{fig:1} 
\end{wrapfigure}

The persistent microgravity environment of space effectively negates the perturbing force of gravity in cold-atom experiments and allows for unprecedented interrogation times of quantum gases in free fall, permitting compact science packages, enhanced precision, and new capabilities for quantum sensors and research facilities. Space also provides access to unique environments for studies of spacetime curvature, tests of the universality of free fall, searches for variations of fundamental constants, and probes for local signatures of dark sector physics via precision measurements~\cite{Safronova2018}. Towards these goals, CAL provides a multiuser platform for seminal demonstrations of quantum technologies in space and for exploring questions with AIs and interacting quantum gases in the unique environment of low Earth orbit that are aligned with the Fundamental Physics goals of the Biological and Physical Sciences Division of NASA's Science Mission Directorate, as summarized in the 2023-2032 Decadal Survey~\cite{BPSdecadal2023}.  

CAL scientists demonstrated numerous pathfinding results with the space-based AI, including phase-shear techniques to observe interference fringes after \SI{150}{ms} of free fall and in-situ sensing of AI laser frequency via atomic phase feedback. They also performed experiments to characterize the ISS vibration environment and explored robust AI protocols for microgravity operation~\cite{williams2024pathfinder}. Further, simultaneous interferometry with $^{87}$Rb and $^{41}$K Bose--Einstein condensates revealed expected interference fringes and phase correlations, providing a footing for future equivalence-principle tests with quantum gases and informing strategies for differential acceleration measurements~\cite{Elliott2023Nature}. Recently, differential AI measurements in CAL have explored relative phase shifts due to magnetic-field curvatures, demonstrating correlated readouts and establishing techniques for space magnetometry and gravity gradiometry~\cite{Meister2025arXiv}. These efforts build on extensive quantum engineering campaigns, which have already achieved gases with total expansion energy below 100 picokelvin and transported over millimeter scales with sub-micron precision~\cite{Gaaloul22}. Filled ultracold atomic bubbles were also studied for the first time in CAL, opening a new topology for quantum matter studies and potential sensor architectures~\cite{Carollo2022Nature}. 

CAL’s autonomous, long-duration operations demonstrate that a space-based quantum gas facility can integrate commercial and custom hardware for impactful, microgravity-enabled fundamental science and space-exploration~\cite{Oudrhiri2024}. These pathfinding efforts accelerate maturation toward mission concepts such as quantum gravity gradiometers, equivalence-principle tests with quantum test masses, and networked quantum sensors. CAL is planned to operate through 2028. On-orbit upgrades of the science module and current driver assemblies for the atom chips and MOT coils were recently launched on the NASA Commercial Resupply Mission NG-24, refreshing and expanding CAL’s science capabilities. Upcoming flight investigations include Efimov physics, few-body and many-body dynamics, Feshbach molecule formation, and advanced dual-species AI experiments. These efforts position CAL to continue de-risking key technologies and delivering high-impact science that advances NASA’s fundamental physics portfolio. 

The Cold Atom Lab (CAL) was designed, built, and is managed by JPL with sponsorship from NASA’s Biological and Physical Sciences Division and the ISS Program Office. The author acknowledges contributions from the CAL team, instrument operations, mission operations, and partner principal investigators across the multi-institution consortium.

\subsection{Longitudinal clock spectroscopy of moving Sr atoms for compact optical clocks \\
            Hidetoshi Katori}

Optical lattice clocks have reached fractional uncertainties at and below the $10^{-18}$ level. At this level, near the Earth's surface, a height difference of 1~cm produces a gravitational redshift of comparable magnitude. This opens new opportunities for chronometric leveling with enhanced temporal response \cite{Tanaka2021Geod}, as well as for precision tests of gravity \cite{Takamoto2020GR}. The technical bottleneck, therefore, shifts from accuracy alone to the stability, temporal response, and robustness required for field deployment. In current optical lattice clocks, intermittent interrogation introduces dead time between preparation and readout, so the clock laser, typically stabilized to an ultrastable cavity and used as the local oscillator (LO), is sampled only periodically; consequently, the clock instability averages down as $1/\sqrt{t}$, and aliasing of LO noise \cite{Dick} can overwhelm the quantum projection noise (QPN) that would otherwise be reduced by interrogating large numbers of atoms.

We presented a route to continuous interrogation for Sr optical lattice clocks. Continuous interrogation removes such interruptions and, in principle, enables $1/t$ scaling of frequency stability until other noise sources dominate. To implement this concept with lattice-trapped atoms, ultracold Sr atoms are continuously transported through a moving optical lattice, so that clock spectroscopy is performed during transport. To maintain Lamb--Dicke confinement and avoid Doppler broadening, the clock laser is applied along the transport axis, realizing longitudinal excitation \cite{Katori2021Longitudinal}. In this geometry, because the atoms are illuminated continuously, the interaction region is defined spatially rather than temporally. We achieve this by localizing the excitation region through magnetic-field-induced mixing \cite{Tai06} on bosonic $^{88}$Sr and by reading out the ${}^1S_0$ and ${}^3P_0$ populations with state-selective detection. This architecture provides the basis for continuous measurement and control of the LO.

The key elements of this architecture were demonstrated experimentally. Under magnetically induced clock excitation, we reconstructed the spatial evolution of the atomic state for atoms transported in the moving lattice, showing that clock excitation is localized and monitored in space. We compensated the $\sim 23$~kHz Doppler shift associated with the moving lattice and recovered the clock resonance. For a 12~mm interaction length and an atomic velocity of about 16~mm/s, the transit time is 0.75~s, corresponding to a Fourier-limited linewidth of about 1.15~Hz, consistent with the measured longitudinal Rabi spectrum. Using an LO stabilized by this continuous spectroscopy scheme, we further demonstrated Ramsey interrogation with dark times of up to 4~s in a second optical lattice clock. These results establish continuous longitudinal clock spectroscopy as a promising route to continuous LO stabilization using an optical lattice clock.

Beyond spectroscopy itself, the broader goal is to translate laboratory optical lattice clocks into robust, self-contained instruments. Such clocks would support chronometric leveling, monitoring of crustal deformation and other geophysical signals, and eventually mobile or space-compatible platforms. For these applications, a major challenge is that achieving the best clock stability still requires heavy reliance on ultrastable reference cavities, which are difficult to reconcile with compact, deployable systems. By enabling uninterrupted monitoring and correction of the local oscillator, the present approach can reduce the degree to which clock stability depends on such references. It therefore provides a concrete step toward optical lattice clocks that preserve high stability while relying less strongly on ultrastable reference cavities.

\subsection{New tests of relativity with optical ion clocks \\
            Christian Sanner}

Is our universe perfectly Lorentz-symmetric or can we detect minute violations of this symmetry on which the Standard Model stands?

Optical clock spectroscopy~\cite{Ludlow2015} on neutral atoms and ions can yield stringent bounds for Lorentz violations (LV) in the electron-photon sector~\cite{Mattingly2005}. The precision of such Lorentz symmetry tests has been improved by several orders of magnitude over the last ten years~\cite{Pruttivarasin2015,Sanner2019,Dreissen2022}. All recent LV tests in earth-bound laboratories rely on the earth's rotation and boost for synchronous LV detection. A space-bound implementation of LV measurements makes it possible to access a wider range of effective lock-in detection frequencies, which is highly advantageous for isolating a true LV signal on top of earth-bound noise patterns. We propose a synchronized earth-space LV test with trapped ytterbium ions employing novel hyperfine-transition-based measurement protocols. This scheme can easily reach a hundred times improved precision without requiring ultrastable laser infrastructure.

\begin{figure}[h!]
 \begin{center}
 \includegraphics[width=0.75\textwidth]{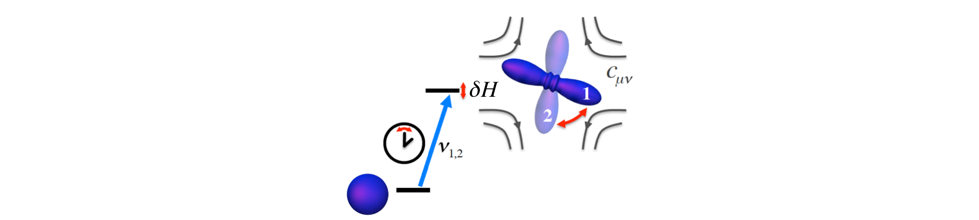}
 \caption{\label{fig:Sanner-fig1}
     Diagnosing spacetime anisotropy through precision spectroscopy of an atomic transition. The clock frequency $\nu$ depends on the orientation of the atomic orbitals with respect to a Lorentz-violating spacetime deformation, which is in the electron sector parametrized via nonzero elements of the $c_{\mu \nu}$ tensor as introduced in the Standard Model Extension (SME) framework~\cite{Colladay1998}.
 }
 \end{center}
 \end{figure}

The platform for the measurements proposed here is an optical atomic clock interrogating singly-charged ytterbium ions. A violation of Lorentz symmetry, i.e., a spacetime anisotropy, will lead to orientation-dependent and boost-dependent clock shifts (see Figure~\ref{fig:Sanner-fig1}).

The electric octupole optical clock transition in Yb$^{+}$ features among all operational atomic clocks the highest sensitivity to LV-induced clock shifts~\cite{Dzuba2016}. Implementing corresponding measurements in a non-earth-bound environment introduces a new modulation timescale and larger rotational boost (e.g., about 90 min for the ISS). By performing synchronized earth-bound measurements one can identify the unique correlation signature of a true LV signal with high fidelity. Because of the complete decoupling this earth-space approach is far superior compared to employing earth-based rotating setups.

In order to simplify the required experimental infrastructure and to achieve the projected hundredfold improvement of the best existing LV bounds, it is desirable to measure the additive frequency modulation $\delta H$ (see Figure~\ref{fig:Sanner-fig1}) against a close-by LV-unaffected reference state. This not only eliminates the requirement for a state-of-the-art ultrastable laser along the optical clock transition, it also makes it possible to exploit the minute-long coherence time demonstrated for microwave spectroscopy~\cite{Fisk1997}.

Beyond the proposed LV measurements and beyond time-keeping and navigation related aspects it is scientifically rewarding to operate a Yb$^{+}$ based atomic clock in space because the $F_{7/2}$ clock states are also highly sensitive dark matter and fine-structure-constant-modulation detectors~\cite{Safronova2018,Flambaum2023}.

\subsection{The space optical clock trade space \\
            David R.~Leibrandt}

Laboratory optical atomic clocks have achieved fractional systematic uncertainties below $10^{-18}$, making them the most accurate measurement devices~\cite{Ludlow2015,Marshall2025,Lindvall2025,Zhang2026,Aeppli2024}.  Space offers longer baselines, larger variations in gravitational potential, and low-disturbance environments, enabling new science.  Applications include planetary and solar science~\cite{Bills2022}, higher-resolution very-long-baseline interferometry (VLBI)~\cite{Gurvits2021,Burt2026,McQuinn2026}, gravitational-wave astronomy in the currently inaccessible $\mu$Hz~\cite{Fedderke2022} and dHz~\cite{Kolkowitz2016,Ganapathy2026} frequency bands, tests of the Einstein equivalence principle~\cite{FOCOS2022}, and searches for ultralight~\cite{Arvanitaki2015,Antypas2022} or topological~\cite{Derevianko2014} dark matter (DM).

For space missions, clock performance must be tailored to the science case and balanced against size, weight, and power (SWaP), cost, and technology readiness level (TRL).  As shown in Fig.~\ref{fig:summary-clocks}, $\mu$m-level ephemerides require $\sim\!10^{-17}$ inaccuracy, while equivalence-principle tests, $\mu$Hz gravitational waves, VLBI up to optical frequencies, and non-common-view comparisons of ground clocks require $\sim\!10^{-18}$ inaccuracy with $\sim\!10^{-15}/\sqrt{\tau}$ instability.  DM searches are limited by clock instability and by the sensitivity of the clock transition to the DM--Standard Model coupling.  Gravitational-wave detection in the dHz band is the outlier, requiring instability $\lesssim\!10^{-18}/\sqrt{\tau}$; a clock with ${1\times10^{-15}/\sqrt{\tau}}$ instability and ${1\times10^{-18}}$ inaccuracy is sufficient for nearly every science case.

Optical clocks are based on locking a laser to a narrow atomic transition and counting its oscillations with a frequency comb~\cite{Ludlow2015}.  Optical lattice clocks interrogate thousands of neutral atoms trapped in a magic-wavelength optical lattice and achieve the best stability~\cite{Aeppli2024}, but need five or six lasers with powers up to $\sim\!1$\,W and a clock laser with $\sim\!10^{-16}$ fractional instability.  Trapped-ion clocks interrogate one or a few ions confined in a radio-frequency Paul trap and have reached the lowest systematic uncertainties~\cite{Marshall2025,Lindvall2025,Zhang2026} with sub-mW laser powers and simpler laser systems, at the cost of higher quantum projection noise.  Key subsystems have flown: the GRACE Follow-On cavity-stabilized laser has operated in orbit since 2018 at $3\times10^{-15}$ instability~\cite{Bachman2017,Abich2019}, frequency combs have flown on sounding rockets~\cite{Lezius2016,Probster2021} and are headed to the ISS~\cite{Kuschewski2024}, and a trapped-ion microwave clock has operated in orbit~\cite{Burt2021}.  Complete optical clocks, however, remain rack-scale or larger laboratory or commercial systems~\cite{TOPTICA2025,Shimadzu2025} at TRL 2 to 3, and a focused technology-maturation program is needed to reach TRL 6.

A single-ion $^{40}$Ca$^+$ optical clock is a compelling near-term candidate because every required wavelength is available directly from a distributed-Bragg-reflector (DBR) diode laser~\cite{Leibrandt2025}: the 729\,nm clock laser, a 732\,nm laser for Doppler cooling on the dipole-forbidden $S_{1/2}\rightarrow D_{3/2}$ transition~\cite{Hendricks2008,Holliman2023}, and 854 and 866\,nm repumpers, each below 1\,mW and referenced through waveguide doublers to an Er:fiber comb and a $10^{-15}$ cavity.  No modulators are required; beams can be switched with shutters or fiber switches.  Compact, laser-machined monolithic ion traps with the well-controlled electric and magnetic fields needed for $10^{-18}$ inaccuracy have already been developed~\cite{Leibrandt2023}.  Instability at the $10^{-15}/\sqrt{\tau}$ level and $10^{-18}$ inaccuracy in a 20\,L, 20\,kg, 100\,W package could be flight ready in about five years (Table~\ref{tab:Leibrandt-trade-space}).  On a longer timescale, photonic integrated circuits (PICs) developed for trapped-ion quantum computing, including integrated narrow-linewidth lasers~\cite{Gundavarapu2019,Jin2021,Loh2025}, light delivery~\cite{Niffenegger2020,Kwon2024}, and modulators~\cite{Hogle2023}, could enable a lower-SWaP, higher-performance multi-ion optical clock.

A pathfinder mission would be prudent before optical clocks are deployed on expensive science missions, many of which require multiple spacecraft in heliocentric orbits or in orbit about planets other than the Earth.  Suitable destinations include medium Earth orbit or higher geocentric orbits, and the lunar surface, where planned Artemis infrastructure could be used.  Coupled with an optical time-frequency transfer link to Earth~\cite{Caldwell2025,Caldwell2023,shen2021}, this pathfinder could improve the gravitational-redshift test by a few orders of magnitude~\cite{FOCOS2022}, search for ultralight and topological DM from $10^{-8}$ to 1\,Hz through comparisons with ground clocks and from $10^{-3}$ to $10^{3}$\,Hz through atom--cavity comparisons, and retire the risk of every subsystem needed by more ambitious follow-on missions.

\begin{table}[h!]
\centering
\scriptsize
\setlength{\tabcolsep}{2pt}
\renewcommand{\arraystretch}{1.2}
\newcommand{\tradehead}[1]{\parbox[t]{0.22\textwidth}{\raggedright#1}}
\newcommand{\tradecell}[1]{\parbox[t]{0.185\textwidth}{\raggedright#1}}
\begin{tabular}{lllll}
\hline
 & \tradecell{\centering Current single-ion technology\par}
 & \tradecell{\centering Future PIC multi-ion technology\par}
 & \tradecell{\centering Current lattice technology~\cite{FOCOS2022}\par}
 & \tradecell{\centering Future lattice technology\par} \\[2pt]
\hline
\tradehead{Example species}
 & \tradecell{$^{40}$Ca$^+$, $^{88}$Sr$^+$, $^{226}$Ra$^+$}
 & \tradecell{$^{40}$Ca$^+$, $^{226}$Ra$^+$, $^{176}$Lu$^+$}
 & \tradecell{$^{87}$Sr, $^{171}$Yb}
 & \tradecell{$^{87}$Sr, $^{171}$Yb} \\[3pt]
\tradehead{Laser system}
 & \tradecell{2--4 DBR lasers up to 1\,mW; $10^{-15}$ FP cavity; no modulators}
 & \tradecell{4--6 diode lasers up to 10\,mW; $10^{-15}$ FP cavity; PIC development needed}
 & \tradecell{5--6 lasers up to 1\,W; not all direct diode; $10^{-16}$ FP cavity}
 & \tradecell{5--6 lasers up to 1\,W; not all direct diode; $10^{-17}$ FP cavity} \\[3pt]
\tradehead{Size, weight, and power (SWaP)}
 & \tradecell{20\,L, 20\,kg, 100\,W}
 & \tradecell{3\,L, 3\,kg, 100\,W}
 & \tradecell{300\,L, 300\,kg, 1\,kW}
 & \tradecell{300\,L, 300\,kg, 1\,kW} \\[3pt]
\tradehead{Statistical uncertainty}
 & \tradecell{$2\!\times\!10^{-15}/\sqrt{\tau}$}
 & \tradecell{$1\!\times\!10^{-16}/\sqrt{\tau}$}
 & \tradecell{$1\!\times\!10^{-16}/\sqrt{\tau}$}
 & \tradecell{$2\!\times\!10^{-19}/\sqrt{\tau}$?} \\[3pt]
\tradehead{Systematic uncertainty}
 & \tradecell{$10^{-18}$}
 & \tradecell{$10^{-19}$}
 & \tradecell{$10^{-18}$}
 & \tradecell{$10^{-20}$?} \\[3pt]
\tradehead{Dark matter sensitivity}
 & \tradecell{$\mathcal{O}(1)$}
 & \tradecell{$\mathcal{O}(1)$}
 & \tradecell{$\mathcal{O}(0.1)$}
 & \tradecell{$\mathcal{O}(0.1)$} \\[3pt]
\tradehead{Flight-ready estimate}
 & \tradecell{$t_0+5$ years}
 & \tradecell{$t_0+10$ years?}
 & \tradecell{$t_0+15$ years?}
 & \tradecell{$t_0+30$ years?} \\[2pt]
\hline
\end{tabular}
\caption{Space optical clock trade space. Dark matter sensitivity denotes the magnitude of the relativistic fine-structure sensitivity coefficient $K_\alpha$. Here, $t_0$ marks the start of funded technology development. FP: Fabry--P\'{e}rot. Question marks identify speculative projections.}
\label{tab:Leibrandt-trade-space}
\end{table}

\subsection{Testing Quantum Physics on Curved Space-Time with Atomic Clocks \\
            Igor Pikovski}

General relativity has been tested to high precision in numerous terrestrial and space-based experiments \cite{will2014confrontation}, yet its interplay with quantum theory remains largely unexplored experimentally. Tests at this interface fall into two broad categories. The first seeks signatures of quantum features of gravity itself, such as single gravitons \cite{tobar2024detecting} or superpositions of the gravitational field \cite{bose2017spin,marletto2017gravitationally}. The second probes how curved spacetime affects the dynamics of quantum systems \cite{dimopoulos2008general,zych2011quantum,zych2012general}. This latter regime, \textit{quantum theory on curved spacetime}, can reveal novel effects even at low energies that depend on post-Newtonian gravity.

Studying proper-time evolution and its superpositions offers a route to post-Newtonian effects in quantum dynamics \cite{zych2011quantum}. While non-relativistic quantum theory typically treats time as a classical parameter $t$, in relativity the time measured by a clock is its proper time $\tau$. Atomic clocks routinely measure proper time and can now resolve gravitational time dilation over mm-scale height differences \cite{bothwell2022resolving,zheng2023lab}. These experiments demonstrate the necessity of accounting for proper time in quantum dynamics: the evolution of an atomic clock is governed locally by $\hat{U}(\tau)$, rather than by $\hat{U}(t)$.

Proper time can itself acquire quantum features, as it depends on position and momentum, which become operators in quantum theory. The resulting evolution $\hat{U}(\tau(\hat{x},\hat{p}))$ can entangle a clock's internal and motional degrees of freedom, leading to interference modulation \cite{zych2011quantum} or even decoherence for many internal degrees of freedom \cite{pikovski2015universal}. For example, a clock in a superposition of two heights evolves under two different proper times $\tau_1$ and $\tau_2$ in superposition. And when the superposition is recombined, the accumulated proper time difference $\Delta\tau=\tau_2-\tau_1$ results in an interferometric visibility $V=|\langle \hat{U}^{\dagger}(\tau_1)\hat{U}(\tau_2)\rangle|$. For a two-level clock with transition frequency $\omega_c$, initially in an equal superposition of its levels, this yields $V=\sqrt{[1+\cos(\omega_c\Delta\tau)]/2}$. This modulation of interference is the signature of how proper times in superposition affect the dynamics.

\begin{wrapfigure}{r}{0.46\textwidth}
  \centering
  \includegraphics[width=\linewidth]{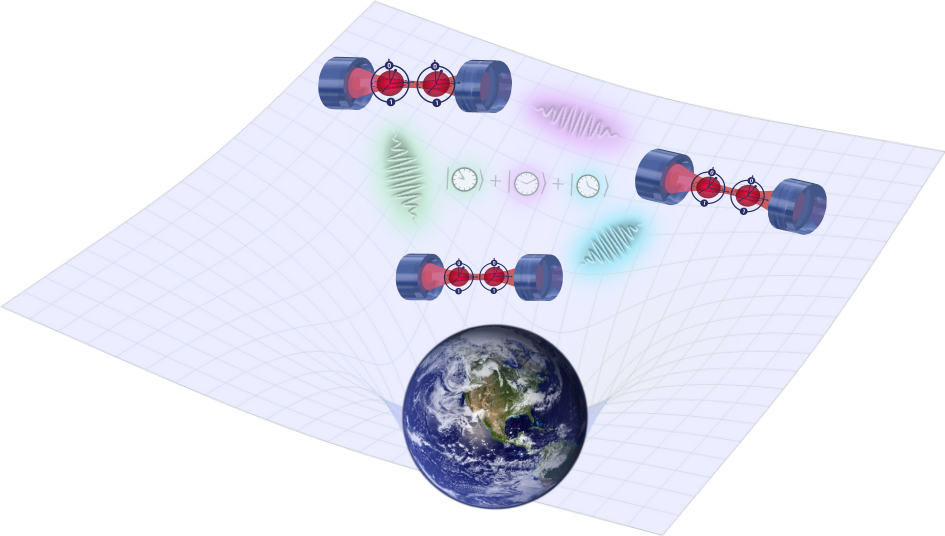}
  \caption{Quantum networks of distributed clocks can probe superpositions of proper times across curved space-time \cite{borregaard2025testing, covey2025probing}, testing the interplay of fundamental principles of quantum theory and general relativity.}
  \label{fig:Pikovski-networks}
\end{wrapfigure}

So far, this signature of superpositions of proper time evolutions has not been observed, as it requires placing clocks in sufficiently large spatial superpositions -- about $10$~m for $1$~s coherence (although the special relativistic version for trapped-ion clocks now seems to be in reach \cite{sorci2025quantum}). It would also be desirable to probe the effect for larger separations, where $\Delta \tau$ is accumulated across curved space-time beyond the linear dependence on height, enabling a genuine test of quantum principles in space-time curvature. There is no empirical evidence whether the principles of quantum theory and those of general relativity remain valid in this domain, or whether one or both have to be modified. But there are theoretical proposals that suggest new physics in this domain, such as modifications to the Born rule \cite{sorkin1994quantum, berglund2023triple} or the unitarity of quantum theory \cite{bassi2017gravitational}.

Quantum networking \cite{komar2014quantum} offers a promising path to achieve proper time interferometry across large distances \cite{barzel2024entanglement, borregaard2025testing, covey2025probing, fromonteil2025non}. It can also be used to probe quantum effects across curved space-time with the following protocol~\cite{covey2025probing}. First one generates entangled W-states of clocks across different locations (Fig. \ref{fig:Pikovski-networks}), which then each evolve according to their own local proper times: $\left| \Psi \right> = \left( \left| c(\tau_1) \right> \left| 0 \right> \left| 0 \right> + e^{i \varphi_2} \left| 0 \right> \left| c(\tau_2) \right> \left| 0 \right> + e^{i \varphi_3}  \left| 0 \right>  \left| 0 \right> \left| c(\tau_3) \right>  \right)/\sqrt{3}$, where each clock state $ \left| c(\tau_i) \right>$ evolves under its local unitary $\hat{U}(\tau_i)$ and $\varphi_i$ are relative phases. Quantum networking techniques can then be used to perform a non-local measurement, such as in the basis of the quantum Fourier transform $\hat{F}_3$. The measurement outcomes will depend directly on the different unitaries: $9\langle \hat{F}_3 \rangle = 3+ 2\langle  \hat{U}^{\dagger}(\tau_1) \hat{U}(\tau_2)\rangle \textrm{cos}(\varphi_{12}) +  2\langle  \hat{U}^{\dagger}(\tau_1) \hat{U}(\tau_3)\rangle \textrm{cos}(\varphi_{13}) +2\langle  \hat{U}^{\dagger}(\tau_2) \hat{U}(\tau_3)\rangle \textrm{cos}(\varphi_{23}) $. The outcomes of such a measurement thus depend on how the different proper time evolutions in the gravitational field are coherently accumulated across the entangled state. For sufficiently large separations, curved space-time would result in the visibility oscillating at several frequencies as the proper time differences $\Delta \tau_{ij}$ are no longer linearly related. This opens the door for first tests of whether curved space-time may alter some of the fundamental principles of quantum theory, such as the Born rule that may generalize to include higher-order interference in quantum gravity \cite{sorkin1994quantum}.

While our current mathematical frameworks are sufficient to calculate the predicted behavior, at some scale the interplay between quantum theory and general relativity may lead to unexpected new effects. Quantum clock interferometry offers a path to test this frontier.

\subsection{Quantum Sensing in Space for Fundamental Physics and Earth Observation \\
            Naceur Gaaloul}

Space-borne quantum technologies, and in particular atom-interferometry-based sensors, are opening a new era in which quantum systems can be exploited as strategic instruments for both fundamental physics and Earth observation. Space offers a unique environment characterized by extended free fall, reduced seismic noise and access to long interrogation times, which are essential ingredients for precision measurements with matter waves. These conditions are highly attractive for applications ranging from time and frequency transfer and inertial sensing to tests of the foundations of physics and the monitoring of the Earth system. At the same time, operating quantum sensors in space imposes demanding constraints on compactness, robustness, power consumption, autonomy and long-term reliability, so that progress requires a close interplay between fundamental atomic physics, engineering, mission design and theory.

Recent years have witnessed important advances in this direction, including pioneering demonstrations of atom optics, quantum gases and interferometric techniques in space-relevant or orbital environments. The realization of Bose--Einstein condensates at picokelvin energy scales in space has established a new platform for ultra-cold-atom experiments under extended microgravity conditions \cite{Gaaloul22}. Further developments have demonstrated quantum gas mixtures and dual-species atom interferometry in space \cite{Elliott2023Nature}, differential atom-interferometric concepts for space magnetometry \cite{Meister2025arXiv}, and apparatus developments targeting quantum-mixture research in microgravity \cite{piest2025apparatus}. Together, these results show that space is no longer only a future environment for quantum technologies, but is already becoming an operational laboratory for quantum gases and matter-wave sensors.

For fundamental physics, one of the most compelling perspectives is provided by the Space-Time Explorer and QUantum Equivalence Principle Space Test, STE-QUEST \cite{gaaloul2025stequest}. This mission concept is designed around a dual-species atom interferometer operating with very long interrogation times in orbit, with the goal of testing the universality of free fall at the level of one part in \(10^{17}\). Such a measurement would probe the weak equivalence principle with quantum objects in a regime inaccessible on ground, while also enabling searches for new physics such as ultra-light dark matter and possible deviations from general relativity. Achieving this level of performance requires a detailed understanding of the spacecraft environment, inertial perturbations, gravity gradients, rotations, wave-front effects and all mission-level systematic errors, as emphasized in recent studies of the platform and environmental requirements for a satellite quantum test of the weak equivalence principle \cite{struckmann2024platform}.

The same technological foundations are also highly relevant for Earth observation. Cold-atom accelerometers and gradiometers offer a measurement principle based on atomic transitions and laser phases, with the potential for excellent long-term stability and reduced instrumental drift. This opens a route toward space quantum gravimetry, where atom-interferometric sensors could contribute to the mapping of temporal variations in the Earth's gravity field. Such measurements are directly connected to the monitoring of climate-relevant mass transport, including continental water storage, ice-sheet evolution, ocean circulation and sea-level change. In this context, the CARIOQA pathfinder mission, recently endorsed in Europe, represents an important step toward the use of atom accelerometers in orbit for future geodesy missions \cite{leveque2023carioqa}. Its goal is not only to demonstrate a new class of quantum sensor in the satellite environment, but also to prepare the methods needed to integrate such instruments into future gravity-field recovery missions.

The progress of the QUANTUS, MAIUS, CUAS, STE-QUEST and CARIOQA activities therefore illustrates a coherent European roadmap for quantum sensing in space. It connects the demonstration of quantum gases and atom optics in microgravity, the development of compact and autonomous cold-atom payloads, the design of ambitious missions for tests of fundamental physics, and the emergence of quantum accelerometry for Earth observation. This roadmap is scientifically powerful because it links questions at very different scales: from the quantum control of dilute atomic gases to the structure of spacetime, the nature of dark matter and the dynamics of the Earth system. Space-borne quantum sensors are thus not merely improved versions of existing instruments; they provide access to qualitatively new measurement regimes at the interface of quantum physics, gravitation, cosmology and geoscience.

\begingroup
\subsection{Optical Ground Station Infrastructure for High-Precision Timing Between Ground and Space \\
            Sascha Schediwy}

Specialised optical ground station infrastructure will be required for the comparison of optical atomic clock timescales between ground and space. The optical ground station is required for precise pointing, acquisition, and tracking of the spacecraft carrying the clock, it needs have a sufficiently large collecting area to receive the spacecraft's timing incoming laser signal on phase-sensitive detectors, efficiently couple this laser light into single mode fibre, transmit a coherent laser timing signal of sufficient optical power at an independently steerable point-ahead angle to hit the spacecraft, be able to remotely access or support a locally hosted ground-based timescale, and ideally be located in an optimal site in terms of environmental and operational conditions.

The Astrophotonics group at University of Western Australia (UWA) has established the TeraNet optical ground station network in Australia (\texttt{www.teranet.space}). This network includes the TeraNet-2 (TN-2) optical ground station located at Geoscience Australia's Yarragadee Geodetic Observatory (YGO) shown in Figure~\ref{fig:Schediwy-tn2}. TN-2 is specifically optimised to support high-precision timing between ground and space.

\begin{figure}[h!]
\centering
\begin{minipage}[t]{0.27\textwidth}
  \centering
  \includegraphics[width=0.82\linewidth]{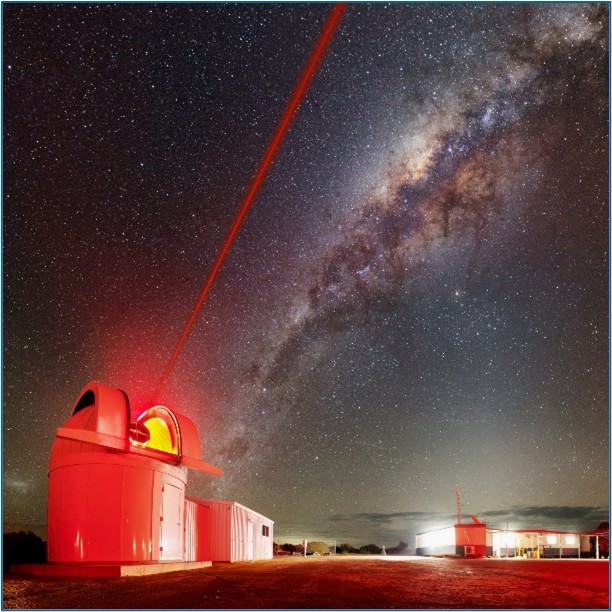}
  \caption{\label{fig:Schediwy-tn2}
      Photo of the TeraNet-2 optical ground station at the Yarragadee Geodetic Observatory.}
\end{minipage}
\hspace{0.04\textwidth}
\begin{minipage}[t]{0.27\textwidth}
  \centering
  \includegraphics[width=0.82\linewidth]{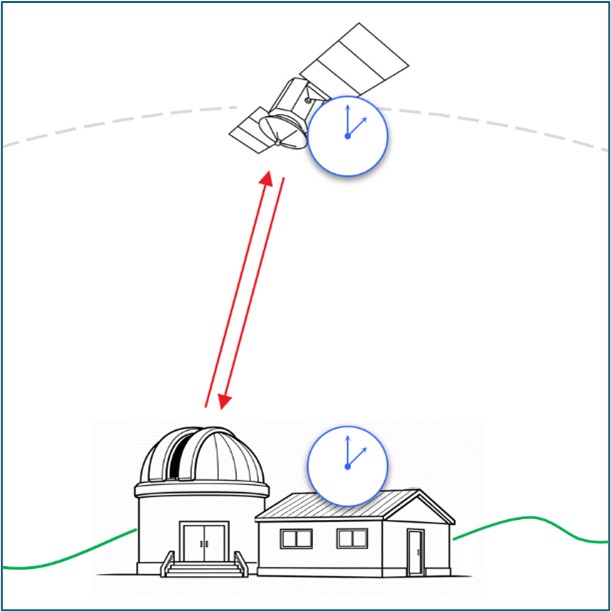}
  \caption{\label{fig:Schediwy-concept}
      Concept for optical atomic clock timescale comparison between ground and space.}
\end{minipage}
\end{figure}

The TN-2 ground station comprises a Planewave Instruments, 70-cm aperture, RC700 telescope, with gold-coated optics that are well-suited for 1550\,nm operations. The telescope is housed within a 4.6\,m two-storey Gambato observatory dome, which is connected to an adjacent building that includes a temperature-controlled laser laboratory and local control room. A HartSCI ClearStar LC adaptive optics system is coupled to one of the telescope's Nasmyth ports to enable efficient single mode fibre coupling. The adaptive optics system includes a separate optical fibre-fed 15-cm transmit telescope with integrated point-ahead capability that includes first-order uplink pointing calibration and correction capability. The other Nasmyth port supports a state-of-the-art, dual-wavelength, kHz-class pulsed satellite laser ranging system, but it also has the provision for a free-space Coud\'{e} path to the adjacent temperature-controlled laser laboratory. The laboratory has an optical table and equipment rack to support future installations of specialised state-of-the-art optical atomic clocks. The laser laboratory includes coaxial cable links to the YGO's local timescale, comprising a suite of geodetic-grade GNSS equipment and a hydrogen-maser atomic clock, and the site is connected via a multi-path AARNet dark fibre network to the UWA campus, ESA's New Norcia Deep Space Tracking Facility, and the Australian Square Kilometre Array telescope site, which operates its own UTC(k).

YGO is one of the world's premier locations for ground-space links. It is a fundamental geodetic observatory that includes the world's most productive satellite laser ranging facility, NASA's MOBLAS-5, a 12-m geodetic VLBI antenna, DORIS transmitter, and geodetic-grade GNSS equipment. The site has a favourable climate with very low annual cloud cover, and especially over the six summer months, and it is centred on a nationally legislated no-fly zone. Considering its low altitude and relative proximity to the coast, it has surprisingly decent astronomical seeing, typically around 1--3 arcseconds. The site is connected to high-speed redundant AARNet internet, grid power, has security controls, and is staffed by Geoscience Australia personnel.

Throughout 2026 and 2027, the Astrophotonics group has plans to use this ground station to support ESA's Atomic Clock Ensemble in Space (ACES) mission through the European Laser Timing experiment, and conduct the first coherent laser link demonstration via reflection off the Ajisai geodetic satellite, utilising the group's velociRAPTOR high-Doppler phase-tracking technology~\cite{DixMatthews2020,McSorley2025} referenced to a commercial-grade portable optical atomic clock. These demonstrations will be used to validate the specific suitability of TN-2 to support future state-of-the-art optical atomic clock fundamental physics space missions, with the experimental concept shown in Figure~\ref{fig:Schediwy-concept}, with TN-2 forming the blueprint for the required deployment of similar specialised optical ground station infrastructure around the world.

Supported by the Australian Space Agency, the Government of Western Australia, and the University of Western Australia.
\endgroup

\subsection{Integrated Visible to Shortwave IR Lasers and Photonics for Cold Atom and Trapped-Ion Quantum Sciences \\
            Daniel J.~Blumenthal}

\begin{figure}[h!]
\centering
\includegraphics[width=0.8\textwidth]{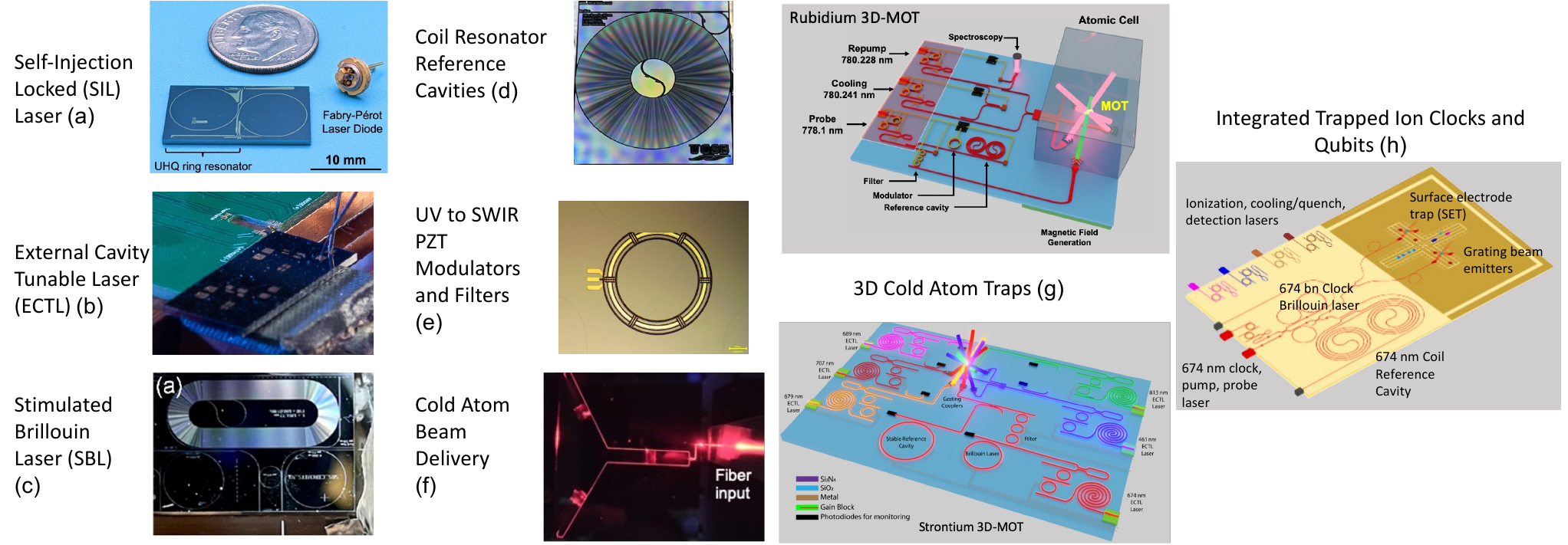}
\caption{Silicon nitride component technologies: (a) self-injection locked (SIL) laser~\cite{Isichenko2024}, (b) extended cavity tunable laser (ECTL)~\cite{Heim2025}, (c) stimulated Brillouin laser (SBL)~\cite{Gundavarapu2019, Liu2024}, (d) coil resonator reference cavities~\cite{Blumenthal2024, Chauhan2026, Heim2025}, (e) PZT optical modulators and filters~\cite{Wang2022, Montifiore2026}, and (f) cold atom beam delivery grating emitters~\cite{Isichenko2023}. Light--matter system-on-chip optical infrastructure: (g) cold rubidium atoms with an integrated 3D magneto-optical trap (MOT) and cold strontium atoms with blue and red MOTs and a lattice~\cite{Blumenthal2024}, and (h) trapped-ion strontium clocks and qubits with a room-temperature surface electrode trap~\cite{Chauhan2026}.
}
\label{fig:LabtoChip}
\label{fig:ChipTechnologies}
\end{figure}

Neutral-atom and trapped-ion quantum sensing and computing today face a situation similar to the tyranny of numbers that limited electronics before the integrated circuit. Quantum experiments rely on complex, lab-scale, bulk-optic laser and optical systems that make up greater than 80\% of the infrastructure used to prepare, control, and measure quantum states, limiting robustness, reliability, cost, and portability. Integrated laser and photonic technologies can now deliver table-top-like performance at a fraction of the cost, volume, and weight, and can bring this infrastructure to the chip scale for reliable space-based light--matter quantum measurements (Fig.~\ref{fig:LabtoChip}).

The photonic integration platform must support a wide variety of atomic wavelengths, laser performance parameters, and light control and delivery functions on the same chip. For atomic transitions from the visible to the shortwave IR (SWIR, 410 nm to 2350 nm), silicon nitride (SiN) is a leading platform~\cite{Blumenthal2018}: it routes light on chip, realizes a wide range of passive components including waveguide-to-free-space beam emitters, and supports lower-speed control modulators based on piezoelectric (PZT) stress-optic actuators~\cite{Blumenthal2018}. Augmenting SiN with other materials adds tunable filtering, high-speed modulation, and frequency shifting; lithium niobate provides these through the linear electro-optic effect and optical nonlinearities. Other platforms extending to the UV include tantalum pentoxide, aluminum nitride (AlN), alumina, lithium niobate, lithium tantalate, and germanium silicate.

Light at the desired wavelengths is generated either directly from a semiconductor laser with the appropriate bandgap or by wavelength translation through nonlinearities such as second-harmonic generation (SHG) and optical parametric oscillation (OPO). Covering 200 nm to 1800 nm requires gain materials including AlN, GaN, InGaN, AlInGaP, GaAs, and InP. No material supports direct emission in the ``green gap'' (500 nm to 600 nm), which is instead reached by SHG and OPO, and watt-level power at the atomic transition remains a future direction. SiN laser designs combine semiconductor gain with passive and tunable on-chip components; examples include extended distributed Bragg reflector (EDBR)~\cite{Huang2019}, extended cavity tunable (ECTL)~\cite{Heim2025}, self-injection locked (SIL)~\cite{Isichenko2024}, and stimulated Brillouin (SBL)~\cite{Liu2024} lasers (Fig.~\ref{fig:LabtoChip}).

Modulation and frequency shifting connect the laser to the atomic transition and prepare and manipulate atomic and qubit states. Thermal, stress-optic, electro-optic, and acousto-optic effects are available depending on the platform. The PZT SiN modulator yields broadband DC to 70 MHz modulation without affecting the waveguide loss or Q while consuming of order 10\,nW~\cite{Wang2022}; it is fabricated at the wafer scale, operates from the visible to the SWIR, and has been realized as phase, Mach--Zehnder, and ring modulators~\cite{Montifiore2026}. For GHz-class modulation, as needed for cooling and repumping, lithium niobate and AlN solutions can be used.

PIC-based beam delivery to the atomic physics package uses waveguide-to-free-space grating-coupled emitters and has been demonstrated for ion trapping~\cite{Niffenegger2020} and rubidium 3D-MOTs~\cite{Isichenko2023}, trapping over 1.3 million rubidium atoms cooled to 200\,$\mu$K. These technologies are at an early stage of readiness for space flight: unpowered SiN chips have been launched to the ISS for radiation testing, powered versions will follow on future missions, and research is focused on integrated atomic clocks and atom interferometers for lunar infrastructure and fundamental physics in space.

\subsection{Towards a community-driven mission roadmap \\
            Oliver Buchmueller}

\begin{figure}[h!]
\begin{center}
\includegraphics[width=0.70\textwidth]{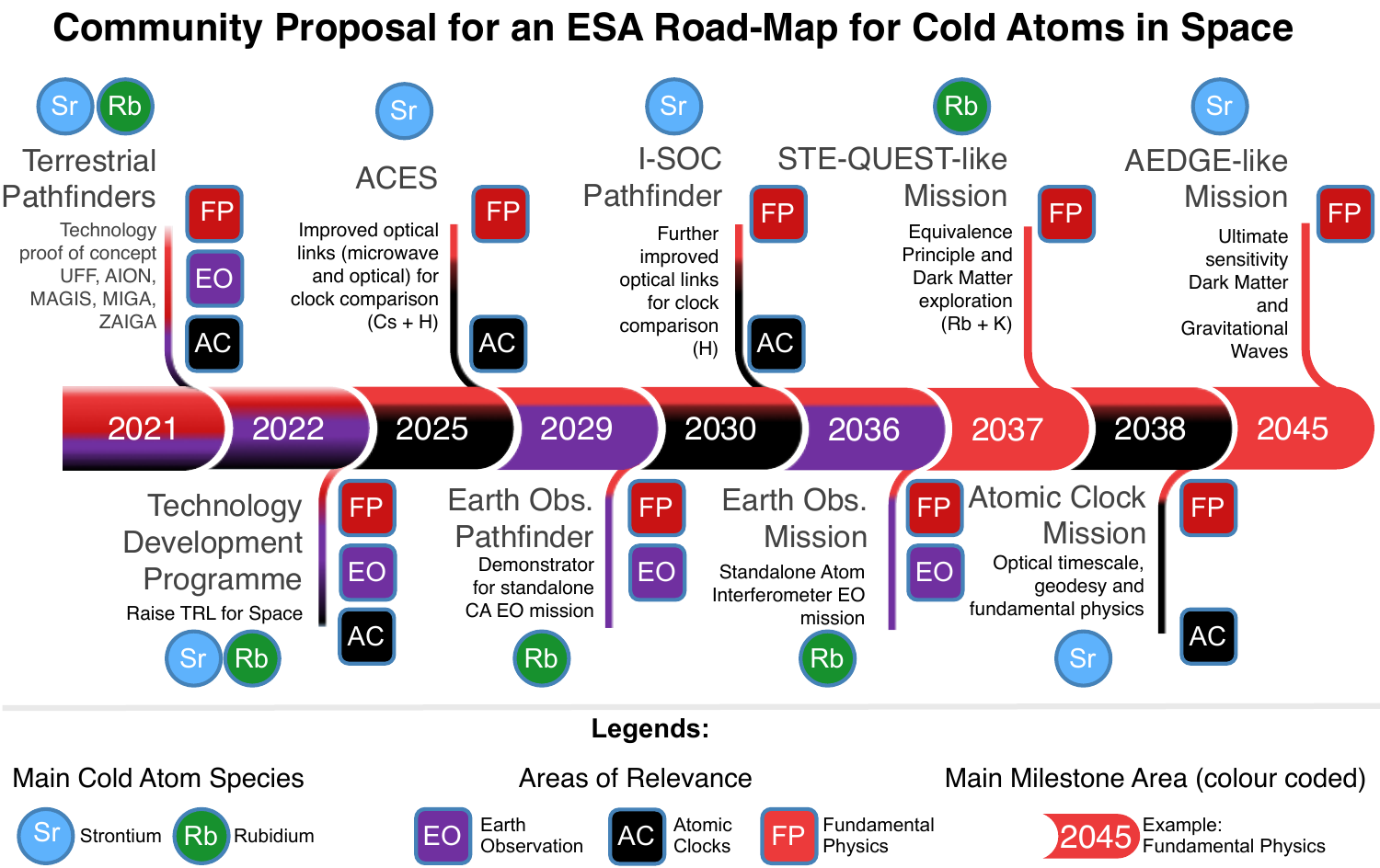}
\caption{\label{fig:roadmap}
Community roadmap for cold atoms in space, as presented in the 2021
Cold Atoms in Space workshop summary~\cite{Alonso:2022oot}. Milestones
are colour-coded by area of relevance: fundamental physics (red), Earth
observation (purple), and atomic clocks (black). The main cold atom
species targeted at each milestone are indicated. Several of the
near-term milestones identified in this roadmap have since advanced
substantially: ACES was installed on the ISS in April 2025, the ESA
Quantum Space Gravimetry Pathfinder (CARIOQA) has entered Phase~A with a
launch target around 2030--2032, and a NASA Quantum Gravity Gradiometer
Pathfinder has been initiated with a similar timeline.
}
\end{center}
\end{figure}

The quantum sensor community has demonstrated that a coordinated,
community-driven approach to mission planning can be both scientifically
productive and effective. A concrete example is provided by the
\emph{Cold Atoms in Space} initiative, which in 2021 organised a virtual
community workshop drawing more than 320 registered participants from 36
countries, spanning the cold atom, astrophysics, cosmology, fundamental
physics, geodesy, and Earth observation
communities~\cite{Alonso:2022oot}. The outcome was a community roadmap
proposing a staged programme of missions and technology development
milestones, submitted to ESA in the context of the Voyage~2050
recommendations and subsequently published as a detailed workshop
summary~\cite{Alonso:2022oot}. The roadmap (Fig.~\ref{fig:roadmap}) illustrates the kind of structured, long-range
planning that the quantum sensor community is capable of producing when
it organises collectively.

Several features of this roadmap process are worth highlighting in the
context of the present workshop. First, the programme was structured
around three interlocking science themes, namely fundamental physics,
Earth observation, and atomic clocks, with explicit synergies identified
between them at every milestone. The cold atom species (strontium,
rubidium) and technology development requirements are mapped onto mission
objectives, providing a clear path from current laboratory capability to
flight readiness. Second, the roadmap was explicitly community-owned
rather than agency-driven: it represented a proposal to ESA for
discussion, not a top-down prescription. This posture proved effective in
engaging both the cold atom technology community and the broader physics
user communities simultaneously.

Several years on from that exercise, the community is well placed to
revisit and update this roadmap in light of significant developments. The
terrestrial pathfinder programme has matured considerably, with AION-10,
MAGIS-100, and VLBAI all advancing towards operation. Several near-term
milestones from the 2021 roadmap have already been executed or initiated,
as noted in Fig.~\ref{fig:roadmap}. At the same time, the science case
has sharpened, particularly for ultralight dark matter and decihertz
gravitational waves as discussed in Sec.~\ref{sec:AI}, and the connection
to multi-messenger astronomy has emerged as a strong additional motivator
for space-based atom interferometry. A similar community-driven process,
adapted to reflect the current landscape and engaging the NASA and DOE
communities alongside ESA, would be a natural and timely next step. The
present Aspen workshop, bringing together the fundamental physics,
quantum sensor, and space instrument communities, provides exactly the
kind of forum in which such an effort could be initiated.

\subsection{PIC development for compact optical clocks \\
            Elena Jordan, Guochun Du, Carl-Frederik Grimpe, Fatemeh Salahshoori, Markus Kromrey, Atasi Chatterjee, and Tanja E. Mehlst\"aubler}

Optical atomic clocks can be applied for timing and navigation, earth observation by relativistic geodesy, as well as for tests of fundamental physics. 
Optical clocks based on a single ion achieve low instabilities, but are eventually limited by quantum projection noise. To overcome this limitation, multiple clock ions can be used in a multi-ion clock. Multi-ion clocks have demonstrated that a reduction in quantum projection noise and shorter averaging time is possible \cite{Tan2019, Cui2022, Keller2024, Filzinger2026}. The setups of these optical clocks are currently too bulky and delicate for space applications.

Photonic integrated circuits (PICs) can significantly reduce the size weight and power consumption (SWaP) of optical clocks and at the same time reduce the sensitivity to vibrations. PIC based components have been employed in lasers for clock interrogation \cite{Loh2025, Chauhan2026} and ion traps with integrated light delivery \cite{Mehta2016, Mehta2020, Niffenegger2020, Ivory2021}. Recently, optical clock operation based on PIC components has been demonstrated \cite{Morrison2025}.  

\begin{figure}[h!]
\centering
\includegraphics[width=0.75\textwidth,trim=0 55bp 0 25bp,clip]{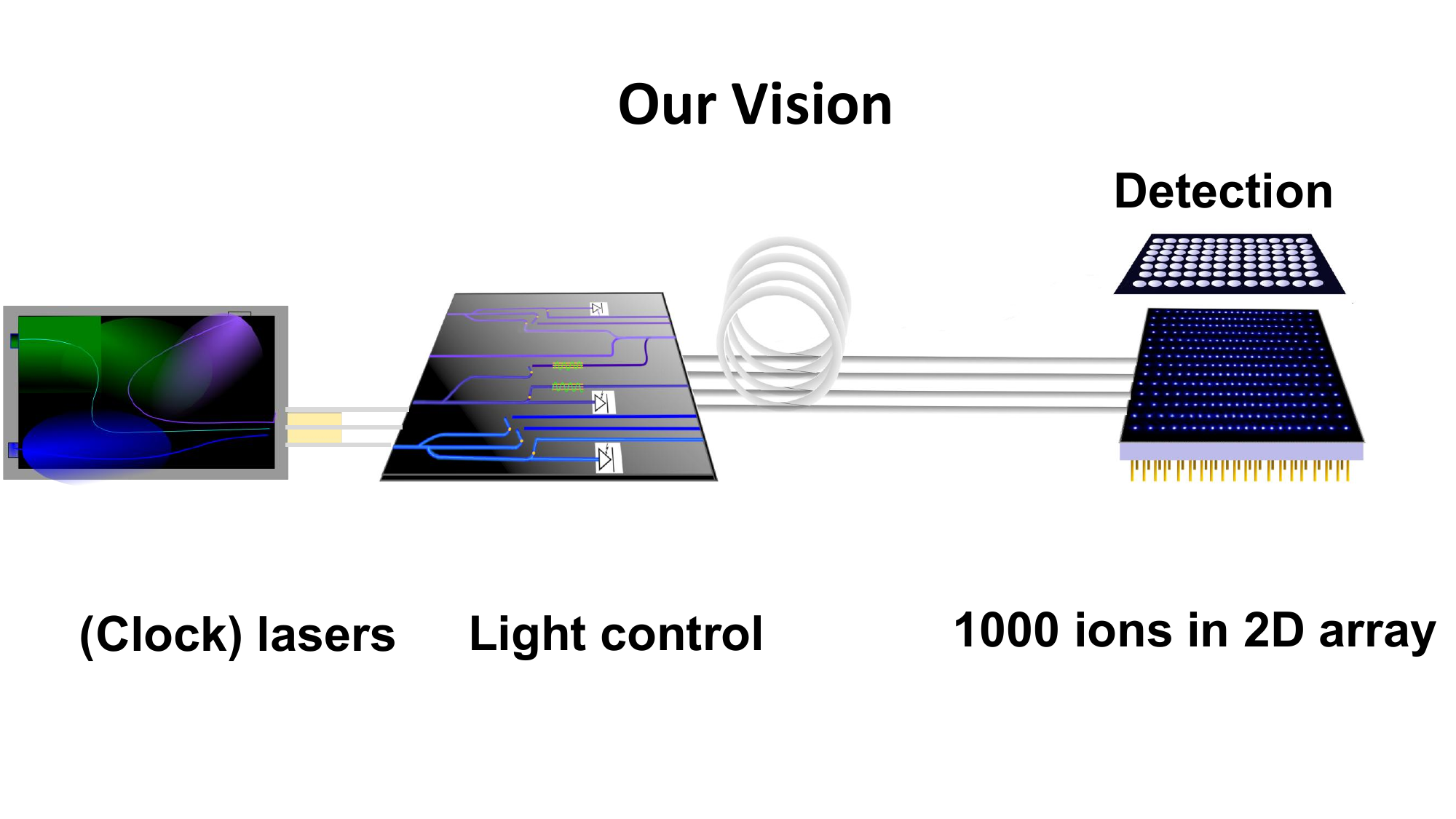}
\caption{\label{fig:Jordan_PICclock}
    Figure on our vision for a clock based on PIC components.}
\end{figure}

In our current projects, we are developing PIC components for a `clock on a chip'. A compact, robust and portable optical clock with low SWaP targeting applications outside the lab, such as earth observation. The components include PIC based lasers for laser cooling, repumping and clock interrogation, a light control unit that contains active components for modulation and switching \cite{Icli2025} as well as passive components for light distribution and combination of multiple wavelengths. The third PIC component is a scalable ion trap for trapping of up to 1000 ions with integrated light delivery for ion addressing and a chip-based detection system \cite {Chatterjee2025} as depicted in Fig. \ref{fig:Jordan_PICclock}. We are currently working with silicon nitride (Si$_3$N$_4$) an established platform for visible wavelengths $>$ 400 nm and aluminum oxide (Al$_2$O$_3$) which has low propagation loss also for wavelengths $<$ 400 nm.  

For addressing the ions with lasers we use Bragg grating couplers on the PIC. The gratings serve for the outcoupling at a well-defined angles and also for beam shaping by using the curvature and varying the effective refractive index across the grating. Using gratings, we generated beam waists between tens of microns and below 5 µm \cite{Grimpe2026, Bannerjee2026}. 

The integration of optics in an ion trap requires apertures in the gold electrodes of the trap that can cause field distortions. We have studied the trapping fields for a surface ion trap using finite element method (FEM) simulations and examined implications for optical clocks. The results show that care should be taken when placing the apertures in the trap design. Another outcome was that transparent conductive coatings are an appropriate mean to minimize field distortions \cite{Du2026}.

Integrated structures also offer the possibility to generate and emit higher order modes, such as a TE01 mode. The outcoupled mode resembles a Hermite Gaussian mode that exhibits a field minimum in the center and at the same time a strong curvature of the field. Together with the inherent stability of monolithically integrated photonics, such structures can be used to excite forbidden clock transitions with reduced AC stark shifts compared to a Gaussian beam \cite{Peshkov2023}.

\begingroup
\setlength{\intextsep}{2pt}
\subsection{Stabilised continuous-wave laser links to support space-based tests of fundamental physics \\
            Benjamin Dix-Matthews}

The prospect of using modern optical atomic clocks as quantum sensors in space creates opportunities for novel tests of fundamental physics~\cite{FOCOS2022}.
However, comparing these clocks with clock-limited performance across ground-space distances is non-trivial~\cite{Caldwell2025}.
Free-space optical frequency transfer is the most promising technique for ground-space clock comparisons; while several significant challenges remain, none appear insurmountable.
Here, we introduce a simple frequency comparison technique and explore these significant challenges.

The technique shown in Fig.~\ref{fig:Dix-Matthews-system} uses counter-propagating continuous-wave optical references to measure the frequency difference between ground and space clocks, $\Delta\nu = \nu_1 - \nu_2$.
The optical references pick up link noise, $\dot{\Delta\phi_{link}^\leftrightarrow}$, as they propagate over the link.
Two heterodyne beats, $\nu^{beat}_G$ and $\nu^{beat}_S$, are measured independently and later combined to cancel out reciprocal link noise and extract the frequency difference between the two clocks.
Non-reciprocal link noise, $\dot{\delta\phi_{link}^{nr}}$, and shot noise, $\dot{\delta\phi_{SN}}$, will remain and are discussed below.

\begin{figure}[h!]
\begin{minipage}[t]{0.6\columnwidth}
  \centering
  \vspace{0pt}
  \scriptsize
  \def\svgwidth{0.85\linewidth}
  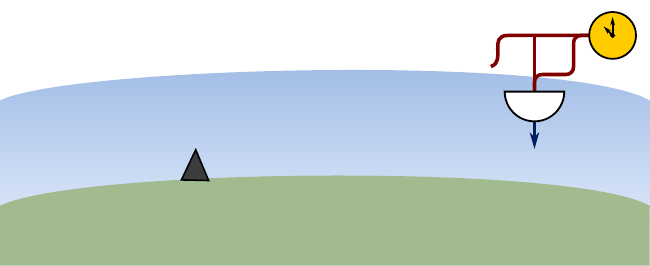
\end{minipage}%
\hfill
\begin{minipage}[t]{0.4\columnwidth}
  \centering
  \vspace{0pt}
  \setlength{\tabcolsep}{1pt}
  \resizebox{0.70\linewidth}{!}{%
  \begin{tabular}{lcc}
     & LEO & GEO \\\hline
    Altitude, $r$ & 500\,km & 35\,786\,km \\
    Transmit $\varnothing$,\ $\omega_{\rm TX}$ & \multicolumn{2}{c}{10\,cm} \\
    Receive $\varnothing$,\ $D_{\rm RX}$ & \multicolumn{2}{c}{70\,cm} \\
    Wavelength, $\lambda$ & \multicolumn{2}{c}{1550\,nm} \\\hline
    $G_{\rm TX}\!=\!(\pi\omega_{\rm TX}/\lambda)^2$ & \multicolumn{2}{c}{$+$106\,dB}\\
    $G_{\rm RX}\!=\!(\pi D_{\rm RX}/\lambda)^2$ & \multicolumn{2}{c}{$+$123\,dB} \\ 
    $L_{\rm fs}\!=\!(4\pi r/\lambda)^{-2}$ & $-252$\,dB & $-289$\,dB \\\hline
    \textbf{Geometric loss} & $\mathbf{-23}$\,\textbf{dB} & $\mathbf{-60}$\,\textbf{dB} \\
    Other losses & \multicolumn{2}{c}{$-20$\,dB} \\
    Transmit power & \multicolumn{2}{c}{30\,dBm} \\\hline
    \textbf{Received power}   & $\mathbf{-13}$\,\textbf{dBm} & $\mathbf{-50}$\,\textbf{dBm}
  \end{tabular}
  }
\end{minipage}
\caption{\label{fig:Dix-Matthews-system}
  \textit{Left:} Simple frequency comparison system to measure the difference, $\Delta\nu$, between the ground clock frequency, $\nu_1$, and the space clock frequency, $\nu_2$. The absolute value in each beat note is removed by assuming $\Delta\nu>>\dot{\Delta\phi_{link}^\leftrightarrow}$.
    \textit{Right:} Downlink link budget estimated using antenna gains~\cite{CarrascoCasado2020,Frost2024}.
}
\end{figure}

\textit{The shot noise limit} depends on received power $P_{\rm sig}$, with a PSD of $S_\phi^{\rm shot}(f) = (2\pi\hbar c)/(\lambda P_{\rm sig})~\text{rad}^2\,\text{Hz}^{-1}$~\cite{Francis2014}.
Converting to modified Allan deviation (MDEV)~\cite{Rubiola2023} yields, ${\rm MDEV}(\tau) = 0.195\sqrt{h_2}\tau^{-3/2}$ where $h_2 = \frac{1}{\nu_0^2}\,\frac{2\pi\hbar c}{\lambda P_{\rm sig}}$, and $\nu_0$ is the nominal optical frequency.
The received power in Fig.~\ref{fig:Dix-Matthews-system} is estimated using the space-to-ground link budget for our ground station~\cite{Schediwy2025}, a satellite transmit power of 1\,W, and a 20\,dB estimate of additional losses for coupling and system inefficiencies.
The shot-noise-limited MDEV at 1\,s is $<10^{-20}$ for $-50$\,dBm received power, far exceeding any clocks.
This suggests shot noise is unlikely to be the dominant constraint, provided that the link can be established and maintain single-mode fibre coupling with the estimated receive power.
We have demonstrated this terrestrially over static~\cite{DixMatthews2021} and dynamic links~\cite{Walsh2022,McCann2024}.

\textit{Non-reciprocal link noise}, $\dot{\delta\phi_{link}^{nr}}$, will not be cancelled, setting a fundamental floor to the performance of the frequency comparison.
Point-ahead angle, anisoplanatism, and finite time-of-flight all introduce non-reciprocal atmospheric noise, with previous work predicting a residual instability near $10^{-17}$ at 1\,s integrating down as $\tau^{-1}$~\cite{Robert2016}.
Physically separated transmit and receive beams are common for large-aperture ground-stations, and will introduce non-reciprocal noise due to mechanical, thermal, and atmospheric effects.
The scale of this effect is an open question.

\textit{Large Doppler shifts}, $\dot{\Delta\phi_{link}^\leftrightarrow}\approx\pm 4$~GHz, are expected in LEO-ground links.
These shifts make high-resolution frequency measurements of the beat notes $\nu^{beat}_G$ and $\nu^{beat}_S$ challenging.
As orbital velocities are well known, most of the Doppler shift can be removed \textit{a priori}.
We published a technique using a single electro-optic-modulator (EOM) to simultaneously correct both the uplink and downlink Doppler-shifts, and demonstrated this technique could suppress Doppler shifts up to 19\,MHz off a flying drone (limited by the 15\,m/s maximum speed of the drone) and achieve $7\!\times\!10^{-18}$ stability after 5\,s of integration~\cite{McSorley2025}.
There are still technical challenges with extending this technique to LEO-ground links, and this remains an active area of research within our group.

\textit{Link intermittency and cycle slips} are an inherent feature of ground-space links due to shadowing, scintillation, pointing loss, and finite orbital pass times.
The resulting dead time and cycle slips will limit the achievable integration times and degrade the stability of the frequency comparison.
Fault tolerant transfer techniques are required to mitigate these effects, including comb-based optical time transfer~\cite{Caldwell2023}, synthetic-wavelength interferometry~\cite{McSorley2025a,McSorley2026}, and gap-aware frequency comparison methods~\cite{Koke2021}.

Free-space optical time and frequency transfer is the best candidate for future comparisons of space and ground optical clocks, however there are some significant challenges that must first be addressed.

This research was supported by the Australian Research Council’s Centre of Excellence for Engineered Quantum Systems (EQUS, project ID CE170100009), the Australian Space Agency, the Government of Western Australia, and the University of Western Australia.
\endgroup

\subsection{Space-based atomic clock searches for variations of fundamental constants \\
            Andrei Derevianko}

Precision clocks turn variations of fundamental constants (FCs) into observable frequency shifts. In the Standard Model the constants are fixed parameters, while many beyond-the-Standard-Model scenarios promote them to dynamical fields that can vary in space and time. Clock comparisons with different sensitivities to $\alpha$, quark masses, and mass ratios therefore provide a laboratory probe of exotic fields. The talk emphasized four complementary signal morphologies: slow drifts, oscillations from virialized fields, stochastic backgrounds, and transients produced when compact field configurations sweep through a clock network \cite{Derevianko2014,Arvanitaki2015,Dailey2020}.

The essential observable is a fractional clock-frequency excursion, $\delta \nu/\nu$, whose size is set by atomic sensitivity coefficients to FC variations and by the couplings of the putative field to Standard Model sectors. In this language, a clock is both a precision oscillator and a calibrated transducer from a change in FCs to a data stream. A single unexplained clock jump is rarely persuasive, but a coherent pattern across many clocks, with the correct propagation speed and geometry, carries much more information: it can encode the size, velocity, direction, and interaction strength of a passing exotic field.

The Global Positioning System provides an unusually powerful example of such a network. Its constellation of Rb and Cs microwave clocks, together with terrestrial receivers and hydrogen masers, forms a roughly $50,000\,\mathrm{km}$ aperture sensor whose public archival data span decades. GPS was built for navigation, but its broadcast microwave signals are tied directly to satellite clock outputs. By measuring carrier phases on the downlinks and combining data from many receivers, one obtains clock, orbit, and receiver-position solutions with timing residuals at the sub-nanosecond level. The resulting data stream, sampled as fast as 30 seconds in standard products, covers many satellites and ground stations over nearly two decades. This is precisely the kind of large, phase-coherent, already-operated instrument that can be repurposed for a rare-event search with a very favorable discovery-potential-to-cost ratio.

A domain wall, monopole, $Q$-ball, or related macroscopic object moving through the constellation at a galactic speed near $300\,\mathrm{km/s}$ would perturb different satellite clocks in a correlated time-ordered pattern. For thin walls, the crossing time over the GPS aperture is of order a few minutes, so the signal appears as a structured sequence of positive and negative frequency excursions across the satellite-clock array rather than as an isolated glitch in one clock. This spatial and temporal pattern is the key discriminator. Template banks over wall direction, speed, and encounter time can search the clock data for the characteristic red-blue tile pattern, with detection thresholds chosen to suppress false alarms. The analysis then layers on practical sanity checks: signal-to-noise and $\chi^2$ filters, consistency between independent JPL and ESA clock products, non-overlap of candidate events, solar-wind directionality tests, and checks against earthquakes or major human-caused disturbances. The first GPS.DM search found no statistically significant domain-wall events in 16 years of data, but improved bounds on quadratic scalar couplings by orders of magnitude and established a reusable Bayesian framework for transient searches in networks of precision sensors \cite{Roberts2017,Roberts2018}.

The same philosophy extends beyond dark-matter encounters in the Galactic halo. In our GPS.ELF program, gravitational-wave detections act as triggers for delayed bursts of exotic low-mass fields (ELFs) emitted by compact-object mergers \cite{Dailey2020}. Because such fields would be massive, they would lag behind the gravitational wave and could imprint an anti-chirp-like transient in the clock network. Applied to GW170817, GPS clock data first showed excess post-trigger noise coincident with enhanced solar activity \cite{Sen2024}; a recent carrier-phase search found no statistically significant ELF signal and set new bounds on quadratic fine-structure coupling over ELF energies near $10^{-18}$--$10^{-14}\,\mathrm{eV}$ \cite{Sen2026}. Looking forward, optical lattice clocks, entangled networks of Yb clocks, and the emerging $^{229}$Th nuclear clocks offer sharper frequency references, portability, and complementary sensitivity to couplings involving photons, electrons, quarks, and gluons. 

The broader message was that existing and future space-qualified quantum sensors can open a low-cost, archival, and multimessenger channel for fundamental physics, especially when their data are analyzed as correlated networks rather than as isolated instruments.

\subsection{Probing the Matter Distribution of the Inner Solar System with Space-Based Atom Interferometry \\
            Susan Gardner}

Recent studies of the Sun and its immediate environment, from the analysis of 
space-based helioseismology and 
MESSENGER and Parker Solar Probe (PSP) mission
data, 
suggest the existence of unexpected
nonluminous mass in the approximate plane of the 
ecliptic, potentially in gross excess of conventional 
assessments~\cite{Alves:2024qvj,Stenborg2024novel}. 
In particular, a recent comparative analysis of the 
Sun's gravitational quadrupole moment, 
from optical studies
of the Sun's surface, as well as from its impact on Mercury's perihelion precession, shows that the slight mismatch in these quantities could be
explained through the existence of a massive, non-luminous circumsolar disk~\cite{Alves:2024qvj}. 
A study of dust-light scattering using the WISPR 
instrument on the PSP 
also reveals a dust overdensity 
at about $0.12\ \rm AU$ from 
the Sun~\cite{Stenborg2024novel}, and 
dust impact data on the PSP itself
shows that both bound and unbound populations
of dust exist in the immediate vicinity of the
Sun~\cite{Szalay2024PSJsizedistrib}. 
These new findings sit comfortably within  
the little that is currently known about the nonluminous matter distribution 
at sub-Galactic scales~\cite{Bechtol_CF2022arXiv220307354B,Chou_cosmicfrontier_2022arXiv221109978C}, 
though with the astronomical data releases from the {\it Gaia} space telescope, 
new constraints are emerging. For example, a vertical Jeans analysis of
a sample of about 200,000 G dwarfs within 
some 2 kpc from the Sun yields a local 
dark matter (DM) density 
of $0.44 \pm 0.13\, \rm GeV cm^{-3} = 0.0117 \pm 0.0035 \,M_{\odot} pc^{-3}$~\cite{Soedling_localdm_2025MNRAS.542.2987S} 
if the dynamics are in steady state 
--- yet we have also observed in both 
stars~\cite{Widrow_wave_2012ApJ...750L..41W,Yanny_SG_wave2013ApJ...777...91Y,Gardner_Noether_2020ApJ...890..110G,Hinkel_data_2020ApJ...893..105H,
Hinkel_2PCF_2023ApJ...942...41H} 
and gas~\cite{Alves_Rad-wave_2020Natur.578..237A,Konietzka_Rad-wave_2024Natur.628...62K}
at scales of a few hundred pc from the Sun 
that they are not. At solar systems
scales, DM constraints come from studies of the 
planetary ephemerides~\cite{Pitjev:2013sfa}, 
from {\it Gaia} 
observations~\cite{Tanga2023solarsystemGaiaDR3}, 
and from the observations of $\gamma$ 
rays that arise from the interactions of
cosmic rays with small, solar system bodies~\cite{deGaetano_fermilat_2023ApJ...951...13D}. 
The last, in comparison with the JPL 
small-bodies database~\cite{JPL_SSB}, reveals the  existence
of {\it many} more objects of about $1 \,\rm km$ or less in 
radius than had been previously thought~\cite{deGaetano_fermilat_2023ApJ...951...13D}. 
Here we wish to determine this mass distribution through its 
spatiotemporal effects, measured using
atom interferometry (AI). The strongest 
constraint thus far 
on solar-system DM is that 
the DM enclosed within a volume set by Saturn's radius
at 67\% CL contributes less than 
$1.7 \times 10^{-10} \rm M_\odot$ in mass~\cite{Pitjev:2013sfa} 
--- in contrast, the limit on the DM mass assuming a uniform distribution of DM as per the analysis of \cite{Soedling_localdm_2025MNRAS.542.2987S},  
gives a limit of about $5 \times 10^{-15} \, \rm M_\odot$. 

As research into the nature of 
DM has 
progressed, we have come to realize that the
genesis of particle DM 
need not be tied
to the energy scale associated with the 
electroweak-symmetry breaking in the 
Standard Model. Rather, masses as light as $10^{-22}\, \rm eV$
appear to be possible~\cite{Hu:2000ke}, speaking
to number densities that would yield 
wave-like properties~\cite{Schive:2014dra,Hui:2016ltb,Hui:2021tkt} --- here DM of 
less than 1 eV in mass is ultralight DM (ULDM). Although solar system DM 
constraints are relatively weak, different DM effects could make 
it more appreciable, through induced spatiotemporal 
variations. For example, a primordial 
black hole could transit the solar system, modifying 
planetary orbits~\cite{Tran2023PBHdarksolarsystem} or be otherwise detectable~\cite{Cuadrat-Grzylowski_DMclump_2024PhRvD.110f3029C}, or ULDM with 
self-interactions could generate a halo around
a planetary object, as in \cite{Budker2023},
with the DM model parameters modifying the formation
times and halo radii. Some ULDM models naturally
contain multiple DM candidates, as in \cite{Kaplan:2015fuy}, e.g., so that multiple
halos, around different planets, would also be possible. 
Independently, the gravitational focusing
of unbound DM in its transit through 
the solar system can also be significant, and estimated using 
the Schroedinger-Poisson framework~\cite{Mocz:2018ium}, with time variations incurred by 
viewing from 
the Earth's orbit. The
ability of AI to measure distant, time-varying 
gravitational effects is long established. 
The effect of the Earth's tides on
the measurement of the gravitational 
acceleration of atom, with $g$ measured 
absolutely to $\Delta g \approx 3\times 10^{-9}g$, was shown 
in 1999~\cite{Peters_falling_1999Natur.400..849P} --- and an AI is in use on the ISS~\cite{williams2024pathfinder}.

AIs have been constructed to be
highly sensitive gradiometers, and the use of dual atomic species with a common laser beam 
reduces noise effects and give sensitive differential measurements~\cite{Overstreet2018,Asenbaum_EP_PhysRevLett.125.191101}.
Such measurements can also be conducted on fast time scales,
offering a nuanced view of possible, time-varying gravitational effects.
A first step towards a gravitational map of the solar system 
would be to monitor the location of a Lagrange point with time. 
At the $L_1$ Lagrange point of the Sun-Earth system, 
at $\approx 0.01\, \rm AU$ towards the Sun from the Earth, 
a spacecraft an Earth radius from $L_1$ should see a restoring force 
of $\approx 1 \times 10^{-7} g$, so that existing AI technology can readily assess
the effect of additional, spatiotemporal gravitational forces, noting that 
the JPL small bodies database~\cite{JPL_SSB} could be used to build a reference model. 

The solar system may 
well contain grossly more dust, rocks, and dark 
matter than previously supposed (and yet be no 
more than a fraction of a percent of the solar system's total mass!). 
With quantum technology, we should be able to assay 
the mass and mass distribution of these small components, though
separating conventional non-luminous matter from DM may require further work.
A combined program of spacecraft and multi-messenger studies, along with the spatiotemporal AI studies we have noted, should 
allow us to separate 
the two, with the possibility of DM coupling to
SM particles, as probed by quantum sensors (including AIs),
offering new insights and cross-checks.

I thank J. Carlton 
and T. Kovachy for their collaboration regarding 
the DM distribution of the solar system and the use of AI, through
spatiotemporal studies, to probe it, 
and J. Mitchell for discussions of terrestrial spatiotemporal AI signals.  
I thank 
G. Alves, S. Broadwater, P. Machado, and M. Zakeri for their
collaboration on DM near the Sun and the 
U.S. Dept. of Energy, Office of Nuclear Physics, 
under DE-FG02-96ER40989 and a Fermilab URA grant 
for its partial support of that work.

\subsection{Multimessenger Astronomy Beyond the Standard Model: New Window from Quantum Sensors \\
            Jason Arakawa}

\begin{figure}[h!]
\begin{center}
\includegraphics[width=0.75\textwidth]{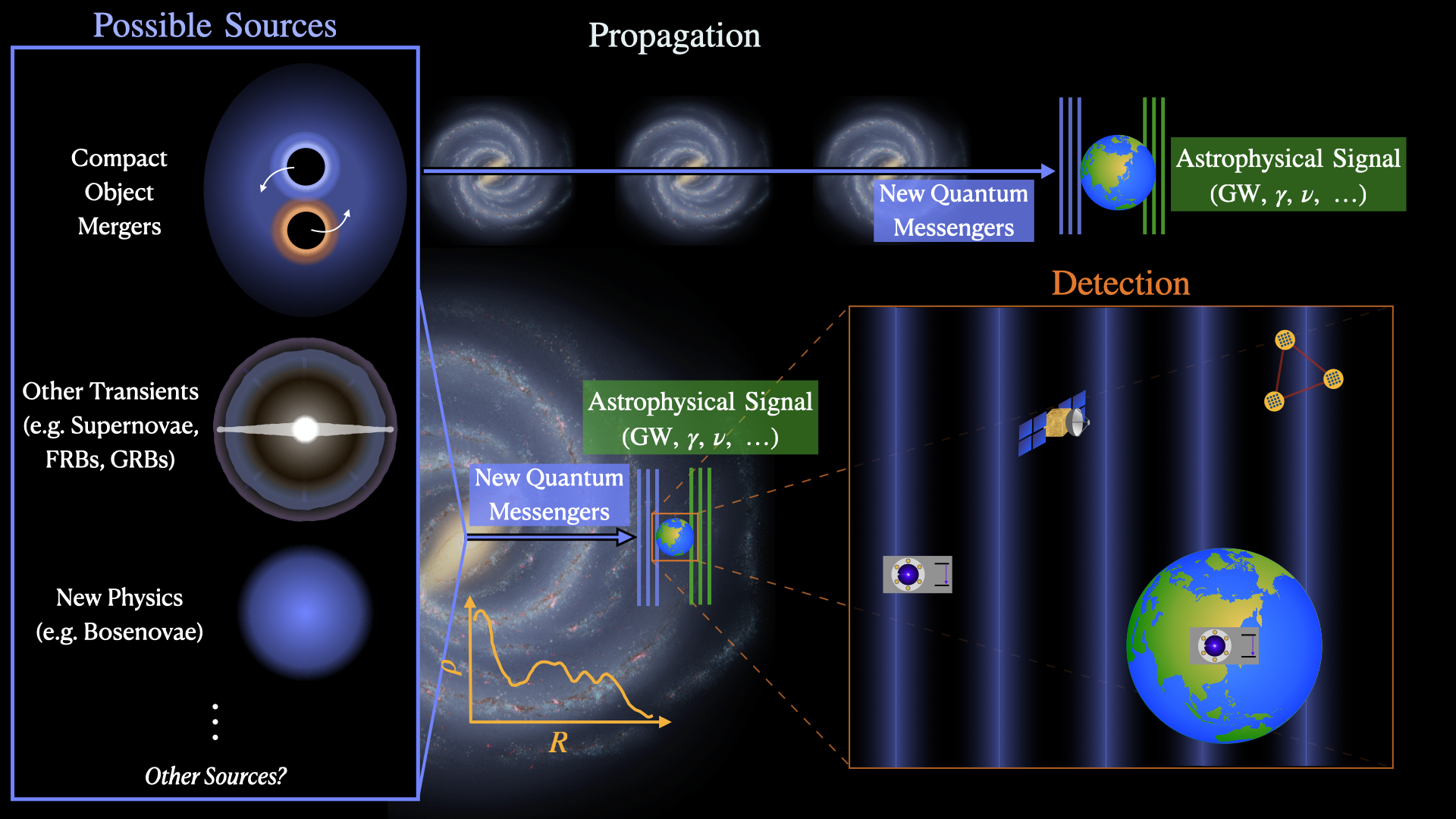}
\caption{\label{fig:Arakawa-MultimessengerAstronomy}
     Schematic of multimessenger astronomy beyond the standard model, with coincident signals of gravitational waves, photons, or neutrinos, with relativistic ultralight fields. We show a non-exhaustive list of possible sources emitting relativistic ultralight fields, the propagation of the fields through normal matter media from the source to the Earth, and subsequent detection by quantum sensors at or nearby the Earth. Figure taken from Ref.~\cite{Arakawa2026}.
}
\end{center}
\end{figure}
The advent of multimessenger astronomy has created a new paradigm of observations, where viewing astrophysical events through multiple fields provides a deeper look into the physics, as gravitational waves (GW), photons, and neutrinos originate from different dynamics within the same event. New ultralight fields $\phi$ may also be produced in large amounts from explosive astrophysical events, such as compact object mergers, supernovae and black hole formation \cite{deGiorgi2024,deGiorgi2026}, or even beyond-Standard Model (BSM) sources \cite{Dailey2020,Eby2021,Arakawa2023,Arakawa2024,Arakawa2026}. This may provide a pathway to multimessenger astronomy with ultralight particles detected by quantum sensors in coincidence with a known physics signal from the same source \cite{Dailey2020, Khamis2024}. Due to their feeble interactions, this program may be one of the only ways to directly detect these particles, especially if they are not associated with dark matter. Additionally, transient signals of this type encode information about microphysics properties that are otherwise difficult to probe \cite{Eby2021}.  

For ultralight fields $\phi$ that interact with Standard Model (SM) fields $i$, e.g. via $\mathcal{L}_{\text{int}} \sim \phi^n \mathcal{L}_{\text{SM},i}$ (for scalar fields) with $n=1,2$, we provide a systematic study of multimessenger astronomy beyond the SM, including treatment of the environmental effects known as screening that can affect both propagation and detection.  
Screening occurs for $n=2$ interactions due to non-zero densities of SM matter in the vicinity of the $\phi$, creating a medium through which $\phi$ is propagating, and can be understood as the SM matter contributing to the effective mass of the $\phi$ \cite{Stadnik2021}. In other words, the SM medium can be treated as a potential barrier for $\phi$ waves, which can both modify $\phi$'s propagation speed or attenuate the $\phi$ signal. If the effect of screening is too large over the course of the propagation, the difference in arrival times (time delay) between the known physics and $\phi$ signals can become too large to correlate the signals \cite{Stadnik2021}. 

The study can be described in three steps: (1) source characterization, (2) propagation, and (3) detection. For the sources, we capture a wide range of possible mechanisms by describing the sources with phenomenological parameters: the total energy emitted in the $\phi$ field $E_{\rm tot}$, the average energy of an individual $\phi$ field $\omega$, the spread in the emitted $\phi$ energies $\delta \omega$, the intrinsic burst duration $t_*$, and the distance from the source to the Earth $R$. We calculate the time delay between GW and $\phi$ signals including screening, and then determine the reach for benchmark experiments, such as the future $^{229}$Th nuclear clock \cite{Campbell2012,Flambaum2020,Peik2021,Higgins2024,Beeks2024}. We find large regions of parameter space across many couplings could be probed by quantum sensors. Much of the parameter space for quadratically coupled scalar fields are subject to critical screening by the Earth, where the Earth and its atmosphere create a strong enough potential barrier that the $\phi$ signals are exponentially suppressed at the Earth's surface. 

Quantum sensors in space would be ideal for this kind of search, as they both avoid screening and benefit from long baselines.

\subsection{New Techniques in Matter-Wave Interferometry for Space-Based Gravitational Wave, Dark Matter, and Fine-Structure Constant Measurements \\
            Jonah Glick}

We present two new techniques for enhancing the performance of space-based atom interferometers.
The first is a quantum optimal control method designed to induce destructive interference among parasitic interferometer paths. These paths are generated by inevitable imperfections in laser pulses and can otherwise limit the sensitivity of an atom interferometer to gravitational waves or dark matter. This new method pushes sensitivities beyond what pulse-level optimization alone can achieve.
The second technique is a method for measuring the atomic recoil velocity using single-photon transitions, with state-selective rather than Doppler selective laser pulses, enabling improved characterization of systematic effects in precision fine-structure constant measurements. We also present a terrestrial Born rule violation search in an atom interferometer as a prototype of a future space-based detector.

\begin{figure}[h!]
\begin{center}
\includegraphics[width=0.8\textwidth]{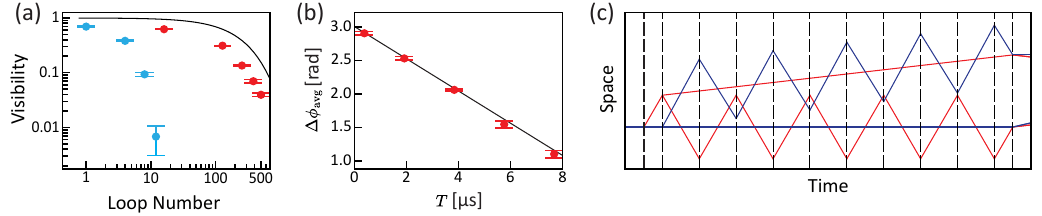}
\caption{\label{fig:Glick-atom_interferometer_responses}
Progress toward multi-loop atom interferometry for space-based sensing of gravitational waves, dark matter, and the fine-structure constant. (a) Enhancement of the interferometer visibility as a function of loop number in a resonant atom interferometer.
Without optimal control (blue points), visibility decays rapidly with loop number; with optimal control (red points), this decay is suppressed, enabling interferometry with 50X more loops and keeping the interferometer visibility within an order one factor of the fundamental spontaneous emission limit (black line).
(b) Recoil-induced phase shift in a $^{88}\text{Sr}$, single-photon atom interferometer as a function of interrogation time $T$. Experimentally measured phase shifts (red dots) agree well with the theoretically predicted dependence of the phase shift on the atom recoil velocity (black line).
(c) Trajectory diagram for a theoretical space-based fine-structure constant measurement under form-factor constraints (see main text). Blue (red) lines denote trajectories in the ground (excited) internal state; dashed lines represent laser pulses.
}
\end{center}
\end{figure}

\textbf{Quantum Optimal Control for Resonant Atom Interferometers:}
An atom gradiometer composed of two atom interferometers separated by a baseline $L$ is sensitive to gravitational waves and ultralight dark matter \cite{Abe2021}, with sensitivity increasing with $L$ and with the number of loops formed by the interferometer arms \cite{Graham2016}. 
Placing an interferometer on each of two satellites in low-Earth-orbit can enable multi-kilometer gradiometer baselines \cite{Hogan2011}.
A central practical challenge in multi-loop (resonant) interferometry is the buildup of parasitic paths from imperfect laser pulses, which can limit the total number of loops that can be performed before the interferometer visibility decays to zero.
We address this with a control scheme focused on tuning the phase of the laser to destructively interfere these paths.
As shown in Fig.~\ref{fig:Glick-atom_interferometer_responses}(a), we experimentally demonstrate a 50X enhancement in the total number of loops achievable in a resonant atom interferometer using this new control scheme \cite{Wang2024}.

\textbf{A New Approach to Precision Measurements of the Fine Structure Constant:}
The fine-structure constant $\alpha$ can be measured through a precision measurement of the velocity with which an atom recoils when it absorbs a photon. Recoil-sensitive atom interferometers have traditionally relied on two-photon transitions, but a greater than $5\sigma$ discrepancy between independent measurements of $\alpha$ \cite{Parker2018, Morel2020} motivates a new measurement based on single-photon transitions \cite{Schelfhout2024}. Unlike two-photon laser pulses, single-photon transitions offer distinct systematic advantages, including the lack of Bloch-oscillation phases.
As shown in Fig.~\ref{fig:Glick-atom_interferometer_responses}(b), we present an experimental measurement of a recoil-dependent phase shift in a single-photon atom interferometer (paper upcoming).
Fig.~\ref{fig:Glick-atom_interferometer_responses}(c) illustrates a theoretical multi-loop extension of the traditional Ramsey-Bordé interferometer sequence which could enable a sensitive space-based $\alpha$ measurement. 
Recoil sensitivity scales with sequence duration and the kinetic energy difference between interferometer arms. Performing interferometry in space enables long sequence durations, but form-factor constraints on the interferometer imposed by the satellite platform can limit the spatial separation of the arms.
The multi-loop configuration enables large kinetic energy differences while keeping spatial separation bounded, and suppresses systematics including those from the Coriolis force \cite{Dickerson2013, Lan2012, Glick2024}.

\textbf{A New Test of Born Rule Violation: }
Certain quantum-gravity theories predict violations of the Born rule which could manifest as an anomalous triple-interference term in a three-path interferometer, beyond the usual pairwise terms \cite{covey2025probing, berglund2023triple, Sinha2010}. We have performed a measurement of this triple interference in a terrestrial three-arm atom interferometer and found no evidence of such a term (paper upcoming).
Future space-based tests, in an environment with varying spacetime curvature, could enable stronger sensitivity to these quantum-gravity effects \cite{covey2025probing}.

\subsection{A High Flux Atom Interferometry Source \\
            Anna L. Marchant}

High flux, cold atomic sources are key in the development of long baseline atom interferometers designed to study gravitational waves and to perform searches for ultra-light dark matter, both on Earth \cite{Badurina2020} and in space \cite{El-Neaj2020}. By increasing the number of atoms in each interferometry sequence, it is possible to enhance the detector performance, limited in part by the atom shot noise, thus pushing to even greater sensitivities. Although many atomic species can be used for interferometry, the presence of a long-lived ‘clock’ state in strontium makes it an ideal candidate for such interferometers as this single-photon transition can be used to perform the interferometry sequence. This single-photon technique allows the same laser beam to address multiple atom sources placed along the baseline, crucially reducing laser phase noise in differential measurements \cite{Graham2013}.  

\begin{figure}[h!]
\begin{center}
\includegraphics[width=0.85\textwidth]{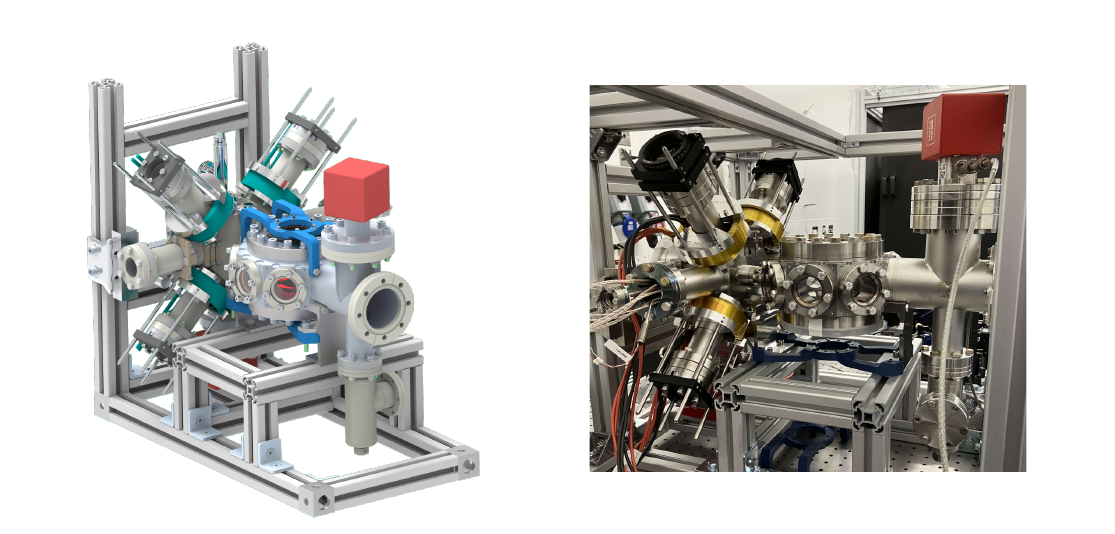}
\caption{\label{fig:Marchant-HiFAIS_fig}
    The HiFAIS system. Left: CAD rendering of the high flux atom interferometry source vacuum chamber showing the source and science chambers, along with a section of the support frame. Right: Photograph of the assembled chamber
}
\end{center}
\end{figure}

The High Flux Atom Interferometry Source (HiFAIS) project has developed a new strontium cold atom source \cite{Walker2026} based on a two-dimensional magneto-optical trap (2D MOT) and a Zeeman slower, shown in Fig.~\ref{fig:Marchant-HiFAIS_fig}. The compact chamber design minimises the distance between the oven, from which strontium atoms effuse through a stainless steel nozzle \cite{Hughes2026}, and the 2D MOT region. Permanent magnets mounted into the chamber body produce the magnetic gradient needed for trapping and contribute to the field profile used for the Zeeman slower. The Zeeman slowing beam enters the chamber though an in-vacuum sapphire window, heated to avoid it becoming coated with strontium metal.  

Characterising the atom flux and velocity distributions from the oven, we find reasonable agreement with models in the free molecular flow regime. We also evaluate several methods to determine the velocity of atoms travelling from the 2D MOT source chamber into a second, science chamber, transferred using a `push' beam. Time-of-flight spectroscopy with a resonant probe beam is used to determine the velocity of the atoms, along with an estimate of the flux capturable by a 3D MOT.  

To fully characterise the `useful' flux from the source, we construct a 3D MOT in the science chamber to measure the loading rate from the 2D MOT. For an oven operating temperature of 465$^\circ$C we measure of loading rate of $10.32(8)~\times~10^9$~atoms/s. This can be further enhanced with the Zeeman slowing beam and the addition of repump light in the 2D MOT, along with frequency modulation of the 2D cooling light. With all of these modifications implemented we achieve a final loading rate of $4~\times~10^{10}$~atoms/s -- to our knowledge, the highest reported loading flux for strontium to date. To assess the lifetime of the vacuum system, we also load atoms into a magnetic trap. Even at the 465$^\circ$C operating temperature, lifetimes of around 8~s are achievable, compatible with the needs of quantum gas experiments. This time can be extended to nearly 24~s by reducing the oven to 400$^\circ$C, and correspondingly the atomic flux to $6.1~\times~10^{9}$~atoms/s. The relatively low operating temperature of the oven, compared to similar systems, also points to potential long-term operation without refilling, a highly desirable feature for large-scale detectors.

\subsection{Gravity Gradiometry with Clock Transitions \\
            Indra Periwal}

Atom interferometry utilizes precisely timed laser pulses to coherently split, redirect, and recombine atomic wave packets. The resulting interference pattern encodes the phase difference accumulated along each path, enabling high-precision inertial measurements. Conventional atom interferometers use alkali atoms driven by two-photon Raman or Bragg transitions. In contrast, clock atom interferometers make use of narrow transitions to metastable states commonly used in atomic clocks. These transitions can be driven resonantly with a single laser beam, which is a promising approach for the suppression of laser frequency noise in long-baseline detectors. To this end, two clouds of atoms are separated by a baseline $L$ and interrogated by common laser pulses in a gradiometer configuration \cite{Graham2013}. This configuration is sensitive to the relative acceleration of the atoms and can be used for gravitational-wave detection. The internal clock of the atom acts as a phase reference, enabling a single-baseline detector. The leading-order phase shift of this configuration scales as $\Delta\phi \propto  h N \omega_a L/c$, where $h$ is the gravitational-wave strain, $N$ is the number of photon recoil kicks corresponding to the large-momentum-transfer (LMT) enhancement factor, $\omega_a$ is the clock transition frequency, and $L/c$ is the light-travel time over the baseline. 

Gradiometer sensitivity scales directly with the interferometer separation $L$, providing the key motivation for longer baselines. The Stanford strontium 10-meter-scale clock gradiometer serves as a prototype for the 100-meter-scale MAGIS-100 detector that is being built in the MINOS shaft at Fermilab. These instruments will probe for ultralight wave-like dark matter and serve as technology demonstrators for an atomic mid-band gravitational-wave detector, bridging the gap between LIGO and LISA. Future extensions to the MAGIS program could be either a terrestrial kilometer-scale detector (MAGIS-km) or a very long-baseline space-based detector (MAGIS-Space). The extremely long baselines in a space based detectors would provide a huge boost in sensitivity without requiring dramatic improvement of the underlying technologies of the sensors. 

The mid-band frequency range bridges a gap in gravitational-wave astronomy, offering a window into massive black hole inspirals, neutron star binaries, and stochastic cosmological backgrounds, while providing valuable pre-merger alerts for multi-messenger observations of LIGO-scale sources. \cite{Hogan2016, Abe2021}.

MAGIS-Space is a proposed space-based detector with a baseline of $L \sim 4 \times 10^7$\,m~\cite{Hogan2016}. The atom interferometers are on two satellites in medium Earth orbit  connected by a heterodyne laser link: a reference laser beam is transmitted between the satellites, and a high-power local oscillator laser near each atom cloud is phase-locked to the incoming wavefront. This architecture decouples the atom-laser interaction strength from the baseline length, since only modest reference beam power is needed to establish the phase lock, while each satellite maintains its own intense local laser to drive atomic transitions. The arrangement preserves the essential common-mode laser-noise cancellation that makes single-baseline operation possible. 

The Stanford strontium 10-meter gradiometer serves as a prototype for these future large-scale detectors. The design leverages strontium's clock characteristics. As an alkaline-earth atom, it possesses a triplet manifold with two clock transitions of complementary linewidths: the intercombination line $^1S_0\to\,^3P_1$ at 689\,nm ($\sim$kHz in $^{88}$Sr) and the ultra-narrow $^1S_0\to\,^3P_0$ clock transition at 698\,nm ($\sim$mHz in $^{87}$Sr). In the 10-meter apparatus, atoms are collected and laser-cooled in a multi-step process. Initial cooling is with a magneto-optical trap (MOT) on the $^1S_0 \to \, ^1P_1$ (461 nm) transitions, where the atoms are cooled to millikelvin temperatures. They are further cooled to microkelvin temperatures with a $^1S_0\to\,^3P_1$ (689 nm) MOT.  The cloud will be then evaporatively cooled in a crossed 1064\,nm optical dipole trap whose depth and curvature are shaped by a reflective spatial light modulator (SLM), reaching nanokelvin temperatures and quantum degeneracy. A matter-wave lensing sequence will collimate the ensemble to picokelvin temperatures for free-fall interrogation. A toss-and-catch double lattice will transport atoms from the cooling chamber into the interferometer region. The atoms will then be launched vertically by an optical lattice. Once the atoms are in free fall, a series of laser pulses will be sent along the vertical axis. To increase the phase sensitivity, we will implement LMT, which increases the effective arm length of the interferometer without physically extending the tower. The resulting matter-wave interference pattern will be imaged. 
 
By demonstrating single-photon clock interferometry, the Stanford strontium 10-meter gradiometer will establish the core sensing architecture for future detectors. The cooling and transport as well as LMT enhancement sequences developed here will be replicated in the MAGIS-100 atom sources. Thus the tower provides a necessary testbed for many of the requirements of MAGIS-100. Ultimately, the 10-meter gradiometer provides the essential proof of concept required to scale toward kilometer-length and space-based gravitational-wave observatories.

\subsection{Towards the rapid generation of Bose-Einstein condensate \\
            Yu-Cheng Tien, Wui Seng Leong, Hsuan-Yu Lin, Liang-Chen Yeh, Yu-Te Tsai, Ko-Wei Cheng, and Shau-Yu Lan}

Quantum gases, such as Bose-Einstein condensates (BECs), have become a powerful platform for exploring many-body physics and implementing quantum sensors. However, traditional preparation relies on evaporative cooling, which depends on the specific collisional properties of the atoms and typically requires seconds to reach degeneracy. This method also suffers from significant atom loss, limiting the overall duty cycle and efficiency. Rapid and efficient preparation of BECs and other quantum gases would improve not only the performance and scalability of precision measurements but also enable new advances in quantum information science.

Recently, a new approach demonstrated the preparation of an unstable $^{85}$Rb BEC with electromagnetically induced transparency (EIT) sideband and adiabatic cooling within a three-dimensional (3D) optical lattice \cite{Xin2024}. This technique led to the quantum regime within 10 ms after the initial sub-Doppler cooling. The EIT cooling process involves two coherent Raman beams coupled to a three-level system in $\Lambda$ configuration. The control beam alters the absorption spectrum of the probe beam, creating a Fano-like profile that strongly suppresses heating transitions during sideband cooling \cite{Morigi2000}. This setup also inherently avoids light-assisted collisions, further preventing atomic loss at high densities.

Between the cooling stages, compression of the atomic cloud is achieved by alternatively switching off one (two) of the lattice beams to form 2D tube (1D pancake) traps. Atoms are allowed to oscillate within these traps along the unrestricted radial directions, then the 3D lattice is restored after one-quarter of the oscillation period ($T/4$) when the spatial distribution is minimized. With lattice beam waists of 60, 60 and 140 $\mu$m, approximately $6 \times 10^5$ atoms are held in the optical lattice after EIT cooling and compression, retaining 95\% of the initial number. In particular, 62\% of the atoms reach the 3D vibrational ground state with a fivefold increase in atomic density, achieving a unity lattice filling. \cite{Xin2024}. In the final stage, the atomic wave functions are merged by adiabatically reducing the lattice potential, while the atoms stay in the ground state. A cross dipole trap holds the atoms in place as the temperature drops, successfully crossing the critical point.

This method is highly versatile, allowing for the production of BEC with a larger number of atoms or at lower temperature. For example, a BEC with $10^9$ atoms could potentially be achieved using a lattice with 1 mm beam waists at the same filling factor. After adiabatic expansion, atoms can also be loaded into a single dipole trap for delta-kick cooling, reaching the ultracold temperatures required for free-space atom interferometers. Furthermore, the short duration within the lattice allows for the use of near-detuned lattice beams with reduced power consumption. Currently, we are working on applying this method to the repulsive isotope $^{87}$Rb to further showcase its potential.

In conclusion, this new approach to quantum gas production can significantly improve the BEC production efficiency with scalable atom numbers. Looking ahead, it can be integrated into a continuous BEC generation scheme, allowing measurements to proceed at one site while new condensates are prepared at another. This would effectively eliminate experimental dead time and maximize throughput, providing an alternative platform for quantum sensor technologies in space and further discoveries in fundamental physics.

\begin{figure}[h!]
 \begin{center}
 \includegraphics[width=0.8\textwidth]{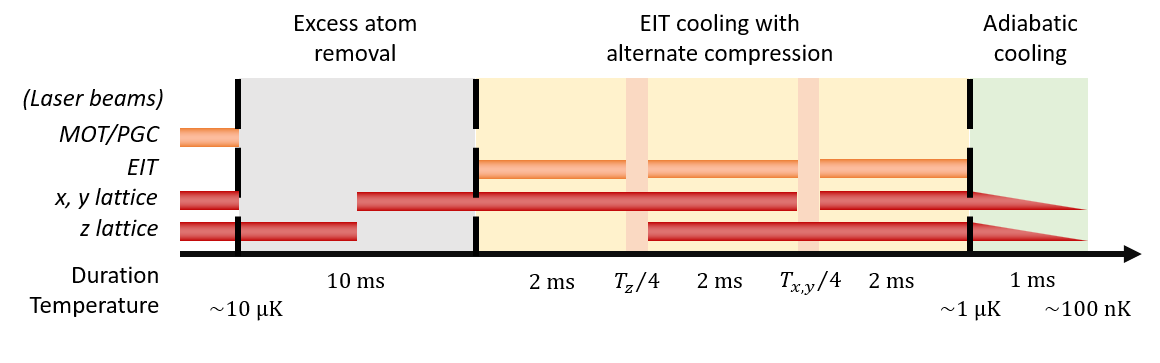}
 \caption{\label{fig:Tien-scheme} Timing sequence for atom preparation (not to scale). $T_x(T_{x,y})$ are the radial oscillation periods in the 2D (1D) traps, each around 0.5 ms. The exact duration and temperature of each step are to be optimized depending on the setup.}
 \end{center}
 \end{figure}

\bibliography{refs}

\end{document}